\documentclass[%
 reprint,
preprintnumbers,
nofootinbib,
 amsmath,amssymb,
 aps,
 authblk,
]{revtex4-2}

\usepackage{graphicx}
\usepackage{dcolumn}
\usepackage{bm}
\usepackage{float}
\usepackage{cancel}
\usepackage{braket}
\usepackage{xcolor}
\usepackage{hyperref}
\usepackage{subfigure}
\usepackage{tabularray}
\usepackage{bm}
\usepackage[bottom]{footmisc}

\begin{document}
\newcommand\myeq{\mathrel{\stackrel{\makebox[0pt]{\mbox{\normalfont\tiny large t}}}{=}}}
\renewcommand{\andname}{\ignorespaces}
\preprint{ADP-23-10/T1219}
\preprint{DESY-23-047}
\preprint{LTH 1336}
\preprint{MIT-CTP/5581}

\title{Constraining beyond the Standard Model nucleon isovector charges}

\author{R. E. Smail$^1$}
\author{M. Batelaan$^1$}
\author{R. Horsley$^2$}
\author{Y. Nakamura$^3$}
\author{H. Perlt$^4$}
\author{\\D. Pleiter$^5$}
\author{P. E. L. Rakow$^6$}
\author{G. Schierholz$^7$}
\author{H. St$\ddot{\text{u}}$ben$^8$}
\author{R. D. Young$^{1,9}$}
\author{J. M. Zanotti$^1$}

\affiliation{$^1$CSSM, Department of Physics, University of Adelaide, Adelaide SA 5005, Australia}
\affiliation{$^2$School of Physics and Astronomy, University of Edinburgh, Edinburgh EH9 3FD, UK}
\affiliation{$^3$RIKEN Center for Computational Science, Kobe, Hyogo 650-0047, Japan}
\affiliation{$^4$Institut f$\ddot{\text{u}}$r Theoretische Physik, Universit$\ddot{\text{a}}$t Leipzig, 04103 Leipzig, Germany}
\affiliation{$^5$PDC Center for High Performance Computing, KTH Royal Institute of Technology, SE-100 44 Stockholm, Sweden}
\affiliation{$^6$Theoretical Physics Division, Department of Mathematical Sciences, University of Liverpool, Liverpool L69 3BX, UK}
\affiliation{$^7$Deutsches Elektronen-Synchrotron DESY, Notkestr. 85, 22607 Hamburg, Germany}
\affiliation{$^8$Universit$\ddot{\text{a}}$t Hamburg, Regionales Rechenzentrum, 20146 Hamburg, Germany}
\affiliation{$^9$Center for Theoretical Physics, Massachusetts Institute of Technology, Cambridge, MA 02139, USA}

\collaboration{QCDSF/UKQCD/CSSM Collaboration}%

\date{\today}
\begin{abstract}
 \vspace{-4mm}
 At the TeV scale, low-energy precision observations of neutron characteristics provide unique probes of novel physics. Precision studies of neutron decay observables are susceptible to beyond the Standard Model (BSM) tensor and scalar interactions, while the neutron electric dipole moment, $d_n$, also has high sensitivity to new BSM CP-violating interactions. To fully utilise the potential of future experimental neutron physics programs, matrix elements of appropriate low-energy effective operators within neutron states must be precisely calculated. We present results from the QCDSF/UKQCD/CSSM collaboration for the isovector charges $g_T,~g_A$ and $g_S$ of the nucleon, $\Sigma$ and $\Xi$ baryons using lattice QCD methods and the Feynman-Hellmann theorem. We use a flavour symmetry breaking method to systematically approach the physical quark mass using ensembles that span five lattice spacings and multiple volumes. We extend this existing flavour breaking expansion to also account for lattice spacing and finite volume effects in order to quantify all systematic uncertainties. Our final estimates of the nucleon isovector charges are $g_T~=~1.010(21)_{\text{stat}}(12)_{\text{sys}},~g_A=1.253(63)_{\text{stat}}(41)_{\text{sys}}$ and $g_S~=~1.08(21)_{\text{stat}}(03)_{\text{sys}}$ renormalised, where appropriate, at $\mu=2~\text{GeV}$ in the $\overline{\text{MS}}$ scheme. 
\end{abstract}

\maketitle

\section{Introduction}\label{sec:intro}
 \vspace{-6mm}
Historically nuclear and neutron beta decays have played an important role in determining the vector-axial (V-A) structure of weak interactions and in shaping the Standard Model (SM). However, more recently, neutron and nuclear $\beta$-decays can also be used to probe the existence of beyond the Standard Model (BSM) tensor and scalar interactions. The interaction of the $W$ boson with the neutron and proton during neutron $\beta$-decay is proportional to the matrix element of flavour changing vector and axial-vector currents between the initial neutron state and final proton state, with coupling constants $g_A/g_V=1.2756(13)$~\cite{Zyla:2020zbs}. It has been identified that the potential existence of BSM tensor and scalar couplings would provide additional contributions to neutron $\beta$-decay~\cite{PhysRev.106.517}. These new BSM contributions are proportional to analogous matrix elements of flavour-changing tensor or scalar operators. To gain sensitivity to these effects the majority of previous and proposed neutron beta decay studies aim to determine one or more of the correlation coefficients included in the differential decay rate for a beam of polarised neutrons \cite{PhysRev.106.517}:
 \vspace{-4mm}
\begin{equation}
\begin{aligned}
\frac{d^3\Gamma}{dE_ed\Omega_e\Omega_\nu}~\propto~& p_eE_e(E_0-E_e)^2\xi\cdot\Big(1+a\frac{\textbf{p}_e.\textbf{p}_\nu}{E_eE_\nu}+b\frac{m_e}{E_e}\\+&\boldsymbol{\sigma}_n\cdot\Big[A\frac{\textbf{p}_e}{E_e}+B\frac{\textbf{p}_\nu}{E_\nu}\Big]\Big),
\label{eqn:decaydist}
\end{aligned}
\end{equation}
where $\boldsymbol{\sigma}_n$ is the neutron spin, $p_e$  is the momentum of the electron and $p_\nu$ is the momentum of the neutrino with energies $E_e$ and $E_\nu$, respectively, and $E_0$ is the end-point energy of the electron. In the SM, $\xi=G_F^2V_{ud}^2(1+3\lambda^2)$, where $\lambda=g_A/g_V$ is the ratio of the axial-vector and vector coupling constants and $G_F$ is the Fermi constant. The neutron decay observables include, $a$, the neutrino-electron correlation coefficient, $b$, the Fierz interference term, $A$, the
beta asymmetry, and $B$, the neutrino asymmetry. Within the SM, the correlation coefficients $a, A$ and $B$ depend solely on the ratio of the axial-vector and vector coupling constants, $\lambda=g_A/g_V$. However the parameter, $b$, is included to account for the case of the hypothetical scalar or tensor couplings in addition to the (V-A) interaction of the SM. Many experiments are underway worldwide with the aim to improve the precision of measurements of these neutron decay observables, two importantly being the neutrino asymmetry $B$ \cite{inproceedings}, and the Fierz interference term $b$ \cite{POCANIC2009211,article1}. The parameter $b$ has linear sensitivity to BSM physics through~\cite{Bhattacharya:2011qm}:
 \vspace{-2mm}
\begin{equation}
\begin{aligned}
b^{\text{BSM}}~=~&\frac{2}{1+3\lambda^2}\Big[g_S\epsilon_S-12\lambda g_T\epsilon_T\Big]\\
\approx~&0.34g_S\epsilon_S-5.22g_T\epsilon_T,
\label{eqn:b}
\end{aligned}
\end{equation}
 \vspace{-4mm}
\begin{equation}
\begin{aligned}
b^{\text{BSM}}_v~=~&\frac{2}{1+3\lambda^2}\Big[g_S\epsilon_S\lambda-4\lambda g_T\epsilon_T(1+2\lambda)\Big]\\
\approx~&0.44g_S\epsilon_S-4.85g_T\epsilon_T,
\label{eqn:bv}
\end{aligned}
\end{equation}
where $\epsilon_T$ and $\epsilon_S$ are the new-physics effective couplings and $g_T$ and $g_S$ are the tensor and scalar nucleon isovector charges. Here $b^{\text{BSM}}_v$ is a correction term to the neutrino asymmetry correlation coefficient, $B$, and $b^{\text{BSM}}$ is an addition to the Fierz interference term $b$ in Eq.~\ref{eqn:decaydist}. Data taken at the Large Hadron Collider (LHC) is currently looking at probing scalar and tensor interactions at the $\lesssim10^{-3}$ level~\cite{ATLAS}. However to fully assess the discovery potential of experiments at the $10^{-3}$ level it is crucial to identify existing constraints on new scalar and tensor operators.\\

Another quantity of interest is the neutron electric dipole moment (EDM), which is a measure for CP violation. In extensions of the Standard Model quarks acquire an EDM through the interaction of the photon with the tensor current \cite{POSPELOV2005119}. The contribution of the quark EDMs, $d_q$, to the EDM of the neutron, $d_n$, is related to the quark tensor charges, $g_T^q$, by \cite{Ellis:1996dg,Bhattacharya:2012bf,PhysRevD.91.074004}:
\begin{equation}
\begin{aligned}
d_n=d_u g^d_T+d_d g^u_T+d_s g^s_T.
\label{eqn:dn}
\end{aligned}
\end{equation}
Here $d_u,~d_d,~d_s,$ are the new effective couplings which contain new CP violating interactions at the TeV scale. The current experimental data gives an upper limit on the neutron EDM of $|d_n| < 1.8\times10^{-26} e$.cm~\cite{PhysRevLett.124.081803}. In calculating the tensor charges and knowing a bound on $d_n$, we are able to constrain the couplings, $d_q$, and hence BSM theories.\\

In recent years there has been an increase in interest from lattice QCD collaborations in calculating the axial, scalar and tensor isovector charges due to their importance in interpreting the results of many experiments and phenomena mediated by weak interactions \cite{PhysRevD.98.034503, Chang:2018uxx, PhysRevD.98.074505, Walker-Loud:2020Lu, PhysRevD.100.034513, PhysRevD.91.054501, PhysRevC.105.065203}.
The QCDSF/UKQCD/CSSM collaborations have an ongoing program investigating various hadronic properties using the Feynman-Hellmann theorem \cite{FH1, FH2, FH3, FH5, Horsley:2012pz, Chambers:2017dov, Can:2020sxc, PhysRevD.105.014502}. Here we extend this work to a dedicated study of the nucleon tensor, scalar and axial charges. We discuss a flavour symmetry breaking method to systematically approach the physical quark mass. We then extend this existing flavour breaking expansion to also account for lattice spacing and finite volume effects to quantify systemic uncertainties. Finally, we look at the potential impact of our results on measurements of the Fierz interference term and the neutron EDM.
\section{Simulation Details}
\begin{figure}[H]
\includegraphics[scale=0.25]{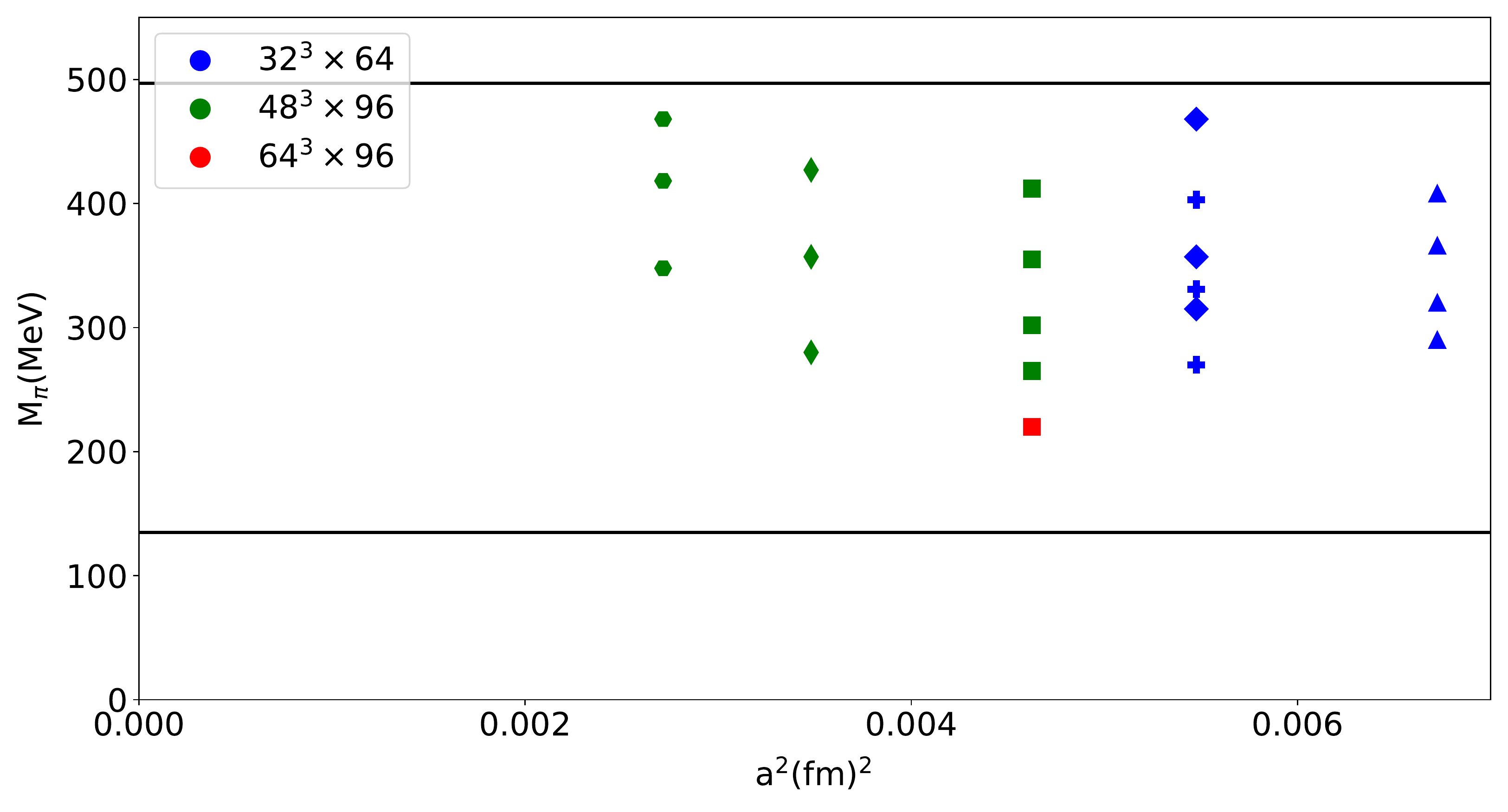}
\caption{\label{fig:config}Lattice ensembles that are used in this work characterised by pion mass, $m_\pi$, and lattice spacing, $a$. The horizontal lines represent the physical pion and kaon masses and the continuum limit occurs as $a\rightarrow0$.}
\end{figure}

For this work we use gauge field configurations that have been generated with $N_f = 2 + 1$ flavours of dynamical fermions, using the tree-level Symanzik improved gluon action and non-perturbatively $\mathcal{O}(a)$ improved Wilson fermions \cite{PhysRevD.79.094507}.
In our simulations, we have kept the bare average quark mass, $\bar{m}=(m_u+m_d+m_s)/3$, held fixed approximately at its physical value, while systematically varying the quark masses around the $SU(3)$ flavour symmetric point, $m_u=m_d=m_s$, to extrapolate results to the physical point \cite{BIETENHOLZ2010436}. We also have degenerate $u$ and $d$ quark masses, $m_u=m_d\equiv m_l$. The coverage of lattice spacings and pion masses is represented graphically in Fig.~\ref{fig:config}.
\begin{ruledtabular}
\begin{table}[H]
\centering
\begin{tabular}{llllll}
 $\beta$& $a$(fm) &Volume & $(\kappa_\text{light},\kappa_\text{strange})$  & $m_\pi$ & $m_K$(MeV) \\ 
\hline
 $5.40$& $0.082$ & $32^3 \times 64$ &$(~0.119930~,~0.119930~)$  & $408$ &$ 408$ \\
 &  &  & $(~0.119989~,~0.119812~)$ & $366$ &$424$ \\
 &  &  & $(~0.120048~,~0.119695~)$ & $320$  & $440$\\
 &  &  & $(~0.120084~,~0.119623~)$ &  $290$ & $450$\\ 
\hline
 $5.50$& $0.074$ & $32^3 \times 64$ & $(~0.120900~,~0.120900~)$ & $468$ & $468$~~*\\
 &  &  & $(~0.121040~,~0.120620~)$ & $357$ & $505$~~* \\
 &  &  & $(~0.121095~,~0.120512~)$&  $315$& $526$~~* \\
\hline
 $5.50$& $0.074$ & $32^3 \times 64$ & $(~0.120950~,~0.120950~)$ & $403$& $403$\\
 &  &  & $(~0.121040~,~0.120770~)$ & $331$ & $435$  \\
 &  &  & $(~0.121099~,~0.120653~)$&  $270$ & $454$\\
 \hline
 $5.65$& $0.068$ & $48^3\times 96$ & $(~0.122005~,~0.122005~)$ &  $412$ & $412$\\
 &  &  & $(~0.122078~,~0.121859~)$ &  $355$ & $441$\\
 &  &  &  $(~0.122130~,~0.121756~)$&  $302$ &$457$ \\
 &  &  & $(~0.122167~,~0.121682~)$ & $265$ & $474$ \\
 &  & $64^3\times 96$ & $(~0.122197~,~0.121623~)$ & $220$ & $485$ \\
\hline
 $5.80$& $0.059$ & $48^3\times 96$ & $(~0.122810~,~0.122810~)$ &  $427$& $427$\\
 &  &  & $(~0.122880~,~0.122670~)$ &  $357$& $456$ \\
 &  &  & $(~0.122940~,~0.122551~)$ & $280$ & $477$ \\ 
\hline
 $5.95$ & $0.052$& $48^3\times 96$ & $(~0.123411~,~0.123558~)$ &  $468$ & $395$\\
 &  &  & $(~0.123460~,~0.123460~)$ & $418$  & $418$\\
 &  &  & $(~0.123523~,~0.123334~)$ & $347$ & $451$
\end{tabular}
\caption{\label{tab:para}Details of lattice ensembles used in this work. * indicates ensembles with a different value of $\bar{m}$, further from the physical $\bar{m}$. The uncertainty on the pseudoscalar masses is between $1$-$3$MeV.}
\end{table}
\end{ruledtabular}
Further information about these ensembles is presented in Table~\ref{tab:para} and Appendix~\ref{appendix2}, Table~\ref{tab:samples}. We have five lattice spacings, $a= 0.082,~0.074,~0.068,~0.059,~0.052$~fm \cite{Bornyakov2015WilsonFA}, enabling an extrapolation to the continuum limit as well as three lattice volumes, $32^3\times 64$, $48^3\times 96$ and $64^3 \times 96$, allowing an extension to the flavour-breaking expansion, which describes the quark mass-dependence of the matrix elements, to also account for lattice spacing and finite volume effects. We also use a bootstrapping resampling technique to compute all statistical uncertainties in our study. \\
 \begin{ruledtabular}
\begin{table}[H]
\centering
\begin{tabular}{l|lll}
 $\beta$&$Z_T^{\overline{MS}}$ & $Z_S^{\overline{MS}}$ & $Z_A$  \\ 
\hline
 $5.40$& $0.9637(23)$ & $0.7034(48)$ &  $0.8671(77)$ \\
 $5.50$ & $0.9644(49)$ & $0.7046(89)$ &  $0.8693(38)$ \\
  $5.65$& $0.9684(54)$ & $0.7153(86)$ &  $0.8754(19)$ \\
  $5.80$& $0.9945(11)$ & $0.6709(23)$ & $0.8913(49)$  \\
  $5.95$& $0.9980(42)$ & $0.6683(94)$ &  $0.8983(43)$
\end{tabular}
\caption{\label{tab:renorm}Renormalisation constants at each value of $\beta$ after chiral and continuum extrapolation across multiple masses with conversion from R$\text{I}^\prime$-MOM to $\overline{\text{MS}}$ at $\mu~=~2$ GeV \cite{Bikerton_2020, Constantinou:2014fka}. }
\end{table}
\end{ruledtabular}
 In order to compare with existing results in the literature we use the renormalisation constants given in Table \ref{tab:renorm}. Table \ref{tab:renorm} summarises the renormalisation constants at each value of $\beta$ after chiral and continuum extrapolation across multiple masses with conversion from R$\text{I}^\prime$-MOM to $\overline{\text{MS}}$ at $\mu~=~2$ GeV. The renormalisation constants are calculated following the method in Ref. \cite{Constantinou:2014fka} and the results first appeared in Ref. \cite{Bikerton_2020}.\\
\section{The Feynman-Hellmann Theorem}
The Feynman-Hellmann (FH) theorem is used to calculate hadronic matrix elements in lattice QCD through modifications to the QCD Lagrangian. The expression for the FH theorem in the context of field theory is \cite{FH1}:
\begin{align}
\frac{\partial E_{H,\lambda}(\vec{p})}{\partial\lambda}=\frac{1}{2E_{H,\lambda}(\vec{p})}\bra{H,\vec{p}}\frac{\partial S}{\partial\lambda}\ket{H,\vec{p}}_\lambda,
\label{eqn:FHFT}
\end{align}
where $S$ is a modified action of our theory so that it depends on some parameter $\lambda$, $S\rightarrow S(\lambda)$ and $E_{H,\lambda}(\vec{p})$ is the energy of a hadron state, $H$. This result relates the derivative of the total energy to the expectation value of the derivative of the action with respect to the same parameter.
\subsection{Application and implementation}
Consider the following modification to the action  of our theory:
\begin{align}
S\rightarrow S+\lambda\mathcal{O}.
\label{eqn:modac}
\end{align}
Then the FH theorem as shown in Eq.~\ref{eqn:FHFT}, provides a relationship between an energy shift and a matrix element of interest:
\begin{align}
\frac{\partial E_{H,\lambda}(\vec{p})}{\partial\lambda}\Big|_{\lambda=0}=\frac{1}{2E_{H}(\vec{p})}\bra{H,\vec{p}}\mathcal{O}\ket{H,\vec{p}}.
\label{eqn:FH5}
\end{align}
Importantly, the right-hand side is the standard matrix element of the operator $\mathcal{O}$ inserted on the hadron, $H$, in the absence of any background field. In lattice calculations, we modify the action in Eq.~\ref{eqn:modac}, then examine the behaviour of hadron energies as the parameter $\lambda$ changes, and finally extract the above matrix element from the slope at $\lambda=0$.\\

In order to calculate the tensor, axial and scalar charges of a baryon, the extra terms we add to the QCD action are:
\begin{align}
S_T&\rightarrow S+\zeta^T_{\mu\nu}\lambda\sum_x \bar{q}(x)\sigma_{\mu\nu}\gamma_5q(x),\label{eqn:FH6}\\
 S_A&\rightarrow S+\zeta^A_{\mu}\lambda\sum_x\bar{q}(x)\gamma_\mu\gamma_5q(x),\label{eqn:FH67}\\
 S_S&\rightarrow S+\lambda\sum_x\bar{q}(x)q(x),\label{eqn:FH7}
\end{align}
where we will take the case of each quark flavour, $q$, separately, $\zeta^T_{\mu\nu}$, $\zeta^A_\mu$ are the phase factors and there are four choices of $\mu$ and $\nu$. The phase factors chosen here are $\zeta^T_{k4}=\zeta^T_{4j}=1$, $\zeta^T_{kj}=i$ and $\zeta^A_{4}=1$, $\zeta^A_{k}=i$. The tensor, axial and scalar charges are related to the baryon matrix elements of the same operators:
\begin{equation}
\begin{aligned}
\bra{\vec{p},\vec{s}}\mathcal{T}_{\mu\nu}\ket{\vec{p},\vec{s}}~=~&-i\frac{2}{m}(s_\mu p_\nu -s_\nu p_\mu)g^q_T,\\
   \bra{\vec{p},\vec{s}}\mathcal{A}_{\mu}\ket{\vec{p},\vec{s}}~=~&2is_\mu g^q_A,\\
   \bra{\vec{p},\vec{s}}\mathcal{S}\ket{\vec{p},\vec{s}}~=~&2m g^q_S,
\label{eqn:ope}
\end{aligned}
\end{equation}
where $\mathcal{T}_{\mu\nu}=\bar{q}\sigma_{\mu\nu}\gamma_5q$, $\mathcal{A}_{\mu}=\bar{q}\gamma_\mu\gamma_5q$ and $\mathcal{S}=\bar{q}q$ \cite{CAPITANI1999548}. In our simulations, we have chosen $\mu=3$, $\nu=4$ and $\vec{p}=0$:
\begin{equation}
  \begin{aligned}
\bra{\vec{0},\vec{s}}\mathcal{T}_{34}\ket{\vec{0},\vec{s}}~=~&2mg^q_T \sigma ,\\
\bra{\vec{0},\vec{s}}\mathcal{A}_{3}\ket{\vec{0},\vec{s}}~=~&2i mg^q_A \sigma,\\
\bra{\vec{0},\vec{s}}\mathcal{S}\ket{\vec{0},\vec{s}}~=~&2m g^q_S,
\label{eqn:op1}
\end{aligned}  
\end{equation}
where, $\sigma=\pm1$, is the spin of the baryon polarised in the $z$ direction.\footnote{Our spin vector is given by $s(\vec{p})=\Big(i\frac{\vec{s}\cdot\vec{p}}{E},\vec{s}(p)\Big)$, where $\vec{s}(\vec{p})=\vec{e} + \frac{\vec{p}\cdot\vec{e}}{m(E+m)}\vec{p}$, with quantisation axis $\vec{n}$ where $\vec{e}=\sigma m \vec{n}$, $\sigma=\pm1$ and $s^2=-m^2$. For our case we have $\mu=3$,~$\nu=4$, $\vec{p}=0$ and $\vec{n}=\vec{e}_3$. Therefore $s_3=\sigma m \vec{e}_3$, $s_4~=~0$ and $p_4=im$.} Hence the FH theorem in Eq.~\ref{eqn:FH5} for the tensor and axial charges gives:
\begin{align}
\frac{\partial E^+_\lambda}{\partial\lambda}\Big|_{\lambda=0}=g^q_{T,A}, && \frac{\partial E^-_\lambda}{\partial\lambda}\Big|_{\lambda=0}=-g^q_{T,A},
\end{align}
where we have dropped the, $H$, subscript as from now on we are only dealing with baryon states and $E^{+/-}$ denotes the baryon energy with spin up/down in the $z$ direction in the presence of a tensor or axial background field (Eq.~\ref{eqn:FH6} and Eq.~\ref{eqn:FH67}) with strength $\lambda$. For small values of $\lambda$, the energy is therefore given by:
\begin{align}
E^{\pm}_\lambda=E_0\pm\lambda g_{T,A}^q +\mathcal{O}(\lambda^3).
\label{eqn:DELTAE}
\end{align}
We have related the change in energy of the hadron state to the spin contribution from the quark flavour $q$. Alternatively, due to the combination of $\pm\lambda$, the spin-down state with positive $\lambda$ is equivalent to the energy shift of the spin-up state with negative $\lambda$. For the scalar we simply have:
\begin{equation}
  \begin{aligned}
\frac{\partial E_\lambda}{\partial\lambda}\Big|_{\lambda=0}~=~&g_S^q,\\
E_\lambda~=~&E_0+\lambda g_S^q +\mathcal{O}(\lambda^2).
\label{eqn:DELTAEscalar}
\end{aligned}  
\end{equation}
Here the insertion is on the quark flavour $q$. For example, we use the perturbed propagator for the $d$-quark in the proton to get the $d$-quark contribution to the nucleon isovector charge. The nucleon isovector charges are then given by the difference between the up and down quark contributions:
\begin{align}
g^{u-d}_{T,A,S}=g^{u}_{T,A,S}-g^{d}_{T,A,S}.
\end{align}
Here we only insert the operator into the propagators used in the quark-line connected contributions; there are no quark-line disconnected terms considered here as they cancel in the case $u-d$. To improve the precision of our results we can take advantage of the fact that we are only interested in energy changes due to changes in $\lambda$, specifically the change in energy around the point $\lambda=0$, with respect to the unperturbed energy. We consider two correlation functions, one calculated at $\lambda=0$ and the other at some finite value of $\lambda$. If we take the ratio of these two quantities, we find:
\begin{equation}
\begin{aligned}
	\frac{C_\lambda(t)}{C(t)}~~&\myeq~~e^{-(E_\lambda-E)t}\frac{E}{E_\lambda}\frac{|A_\lambda|^2}{|A|^2}.
\end{aligned} 
\label{eqn:corr_rat}
\end{equation}
The exponential dependence on $t$ now contains the difference in energies between the unperturbed energy and the energy at some $\lambda$.
$C$ and $C_\lambda$ are both measured on the same configurations, so both will have correlated noise. Using this ratio to determine energy differences has the advantage that the noise will largely cancel, leaving to a more reliable energy shift. We can also constrain our fit function to pass through zero by construction as there is no difference in energies at $\lambda=0$.\\

The extraction of hadron matrix elements in lattice QCD demands careful attention to contamination from excited states. Excited-state contamination has an impact on the study of standard baryon three-point functions due to the presence of weak signal-to-noise behavior at large Euclidean times. Various techniques are used to address excited-state contamination, one of which is the variational method. The variational method has been widely successful in spectroscopy investigations \cite{ALPHA, PhysRevD.82.034505, MAHBUB2012389, PhysRevD.91.094509, PhysRevD.87.094506, PhysRevLett.108.112001, PhysRevD.84.074508}, and has also found application in the analysis of hadronic matrix elements \cite{PhysRevD.92.034513, PhysRevLett.114.132002, PhysRevD.91.074503, OWEN2013217, PhysRevD.78.114508, PhysRevD.93.114506}. Another popular method is the ``two-exponential fit" and ``summation" methods seen in Refs.~\cite{PhysRevD.78.114508, PhysRevD.93.114506, DINTER201189, PhysRevD.89.094502, PhysRevD.92.054511, PhysRevD.90.074507, PhysRevD.86.074502}. A summary of these methods as well as a comparison between them can be seen in Ref.~\cite{Dragos:2016rtx}.
\begin{figure}[H]
\includegraphics[scale=0.28]{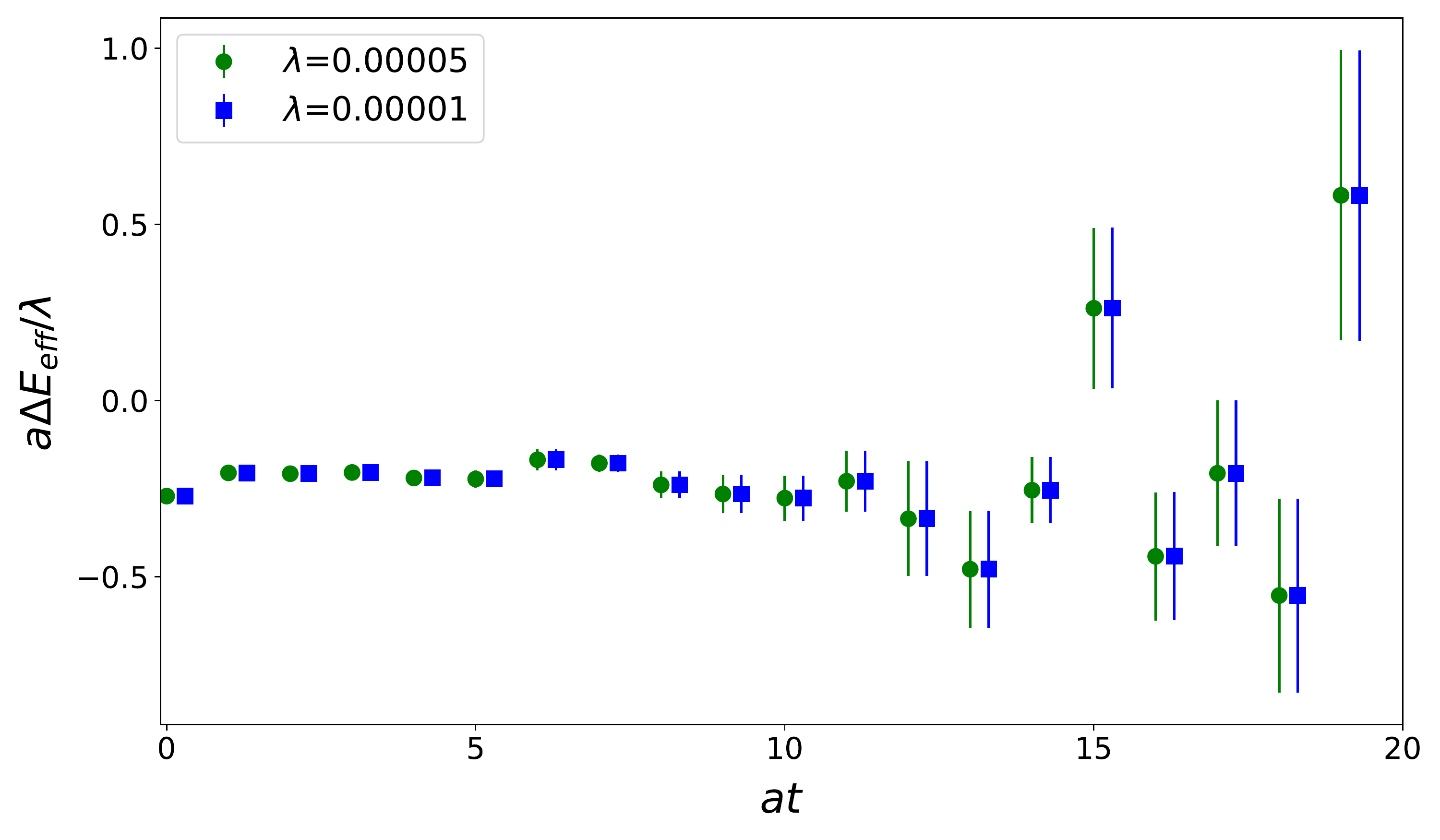}
\caption{\label{fig:FHeffmass}Proton effective mass for the ratio (Eq.~\ref{eqn:corr_rat}) divided by $\lambda$, for the down quark at two different values of $\lambda$, calculated at $a=0.068$fm, $(\kappa_l,~\kappa_s)=(0.122167,~0.121682)$ for the tensor. The points have been offset slightly for clarity.}
\end{figure}Since in this investigation hadron energies are extracted from two-point functions, control of excited state contamination in the Feynman-Hellmann is simplified compared to standard three-point analyses. For example Fig.~\ref{fig:FHeffmass} shows the effective energy shift for the ratio (Eq.~\ref{eqn:corr_rat}) divided by $\lambda$ for the down quark at two different values of $\lambda$. In Fig.~\ref{fig:FHeffmass} we see a plateau in the effective mass indicating a clear region where the ground state can be isolated.
\begin{figure*}
  \centering
  \subfigure[]{\includegraphics[scale=0.23]{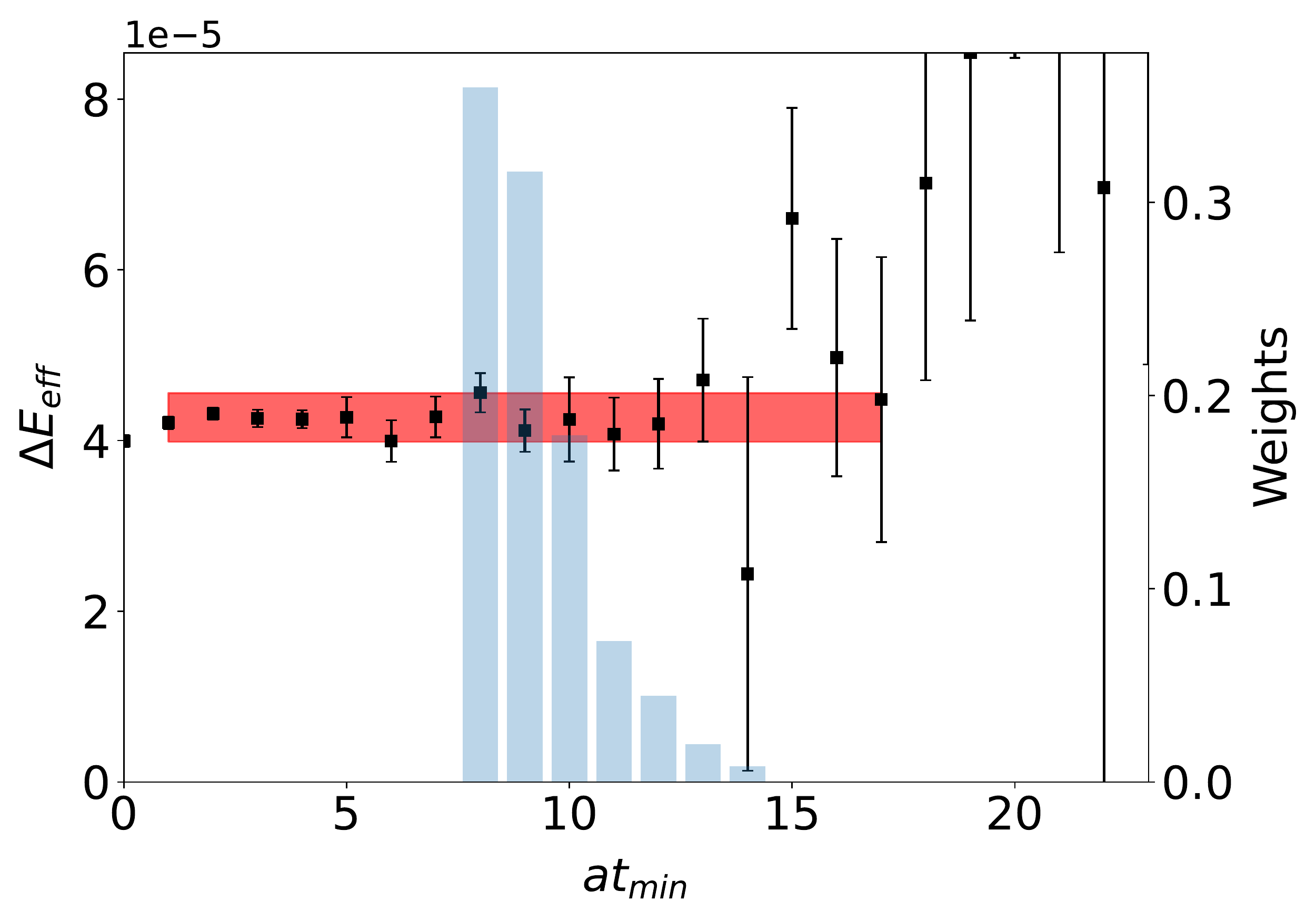}}
  \subfigure[]{\includegraphics[scale=0.23]{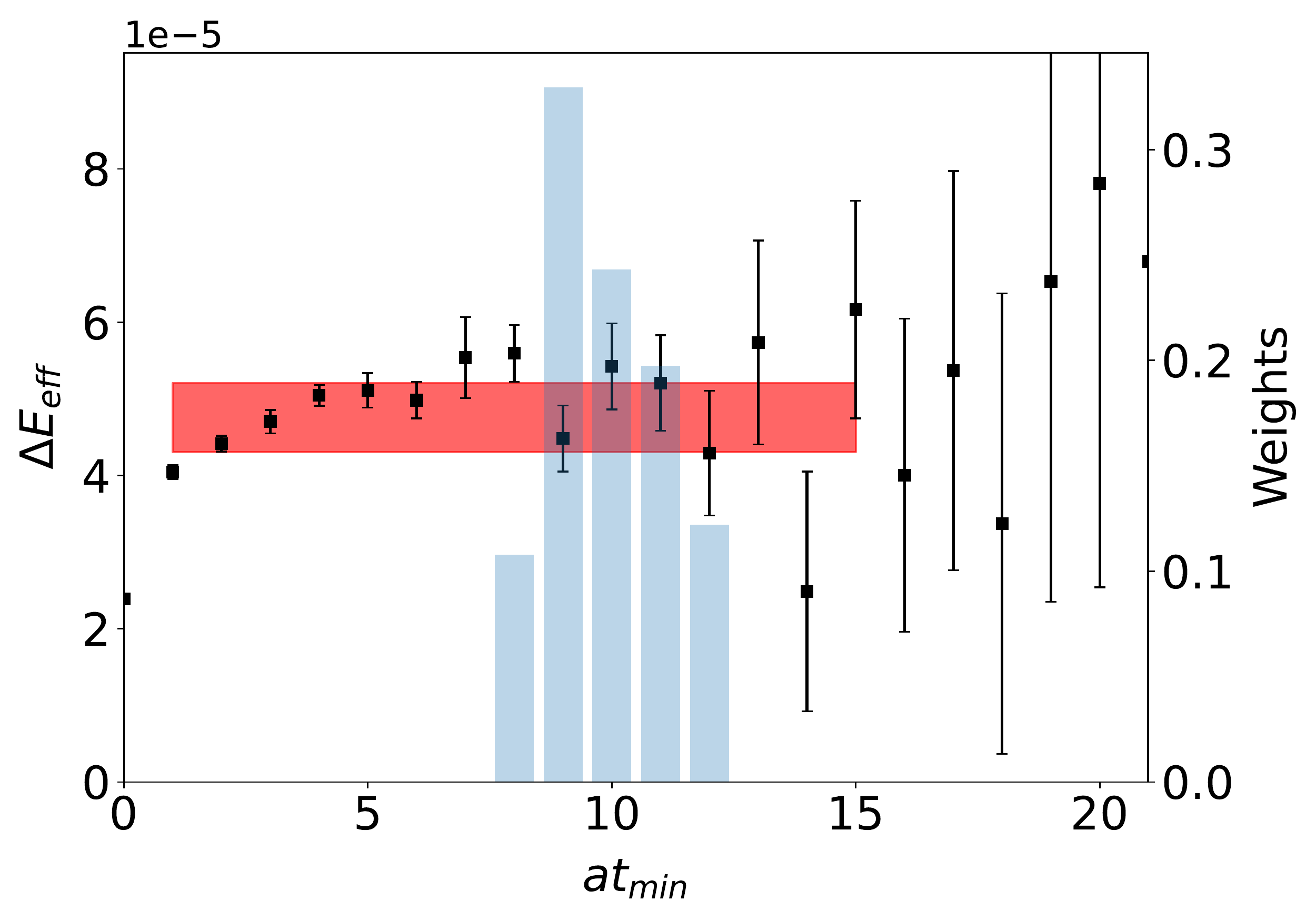}}
    \subfigure[]{\includegraphics[scale=0.23]{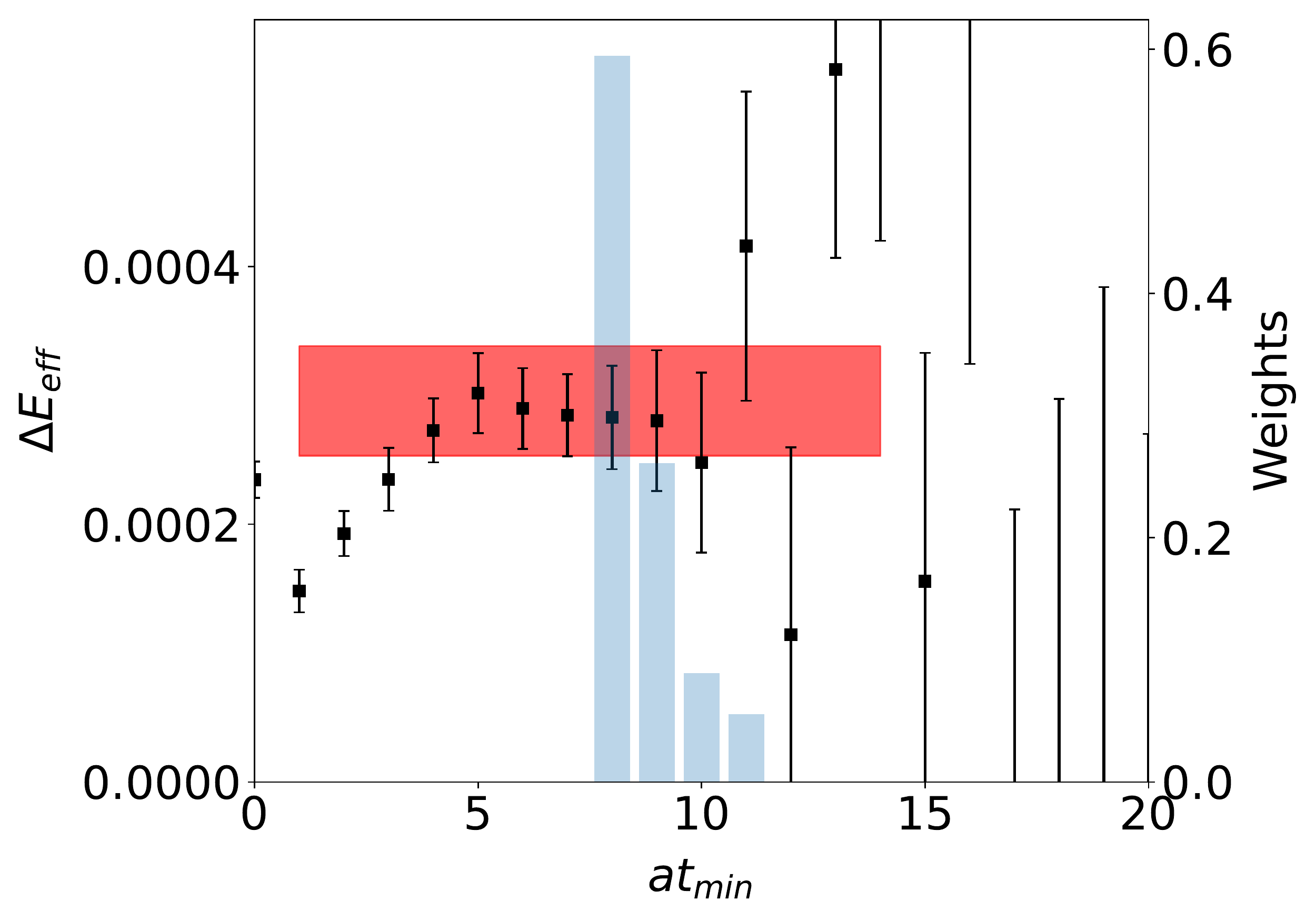}}
\caption{\label{fig:WARES}Proton effective mass for the ratio (Eq.~\ref{eqn:corr_rat}) for the up quark at $\lambda=0.00005$, for spin-down in the tensor (a) and the axial (b) with the scalar results show in (c), calculated at $a = 0.068$fm, $(\kappa_l,~\kappa_s)=(0.122167,~0.121682)$. The blue bar graph shows the weight of each fit result for the value of $t_\text{min}$. The horizontal (red) band is the weighted average value, where the band includes the combined statistical and systematic uncertainty.}
\end{figure*}
\begin{figure*}
  \centering
  \subfigure[]{\includegraphics[scale=0.23]{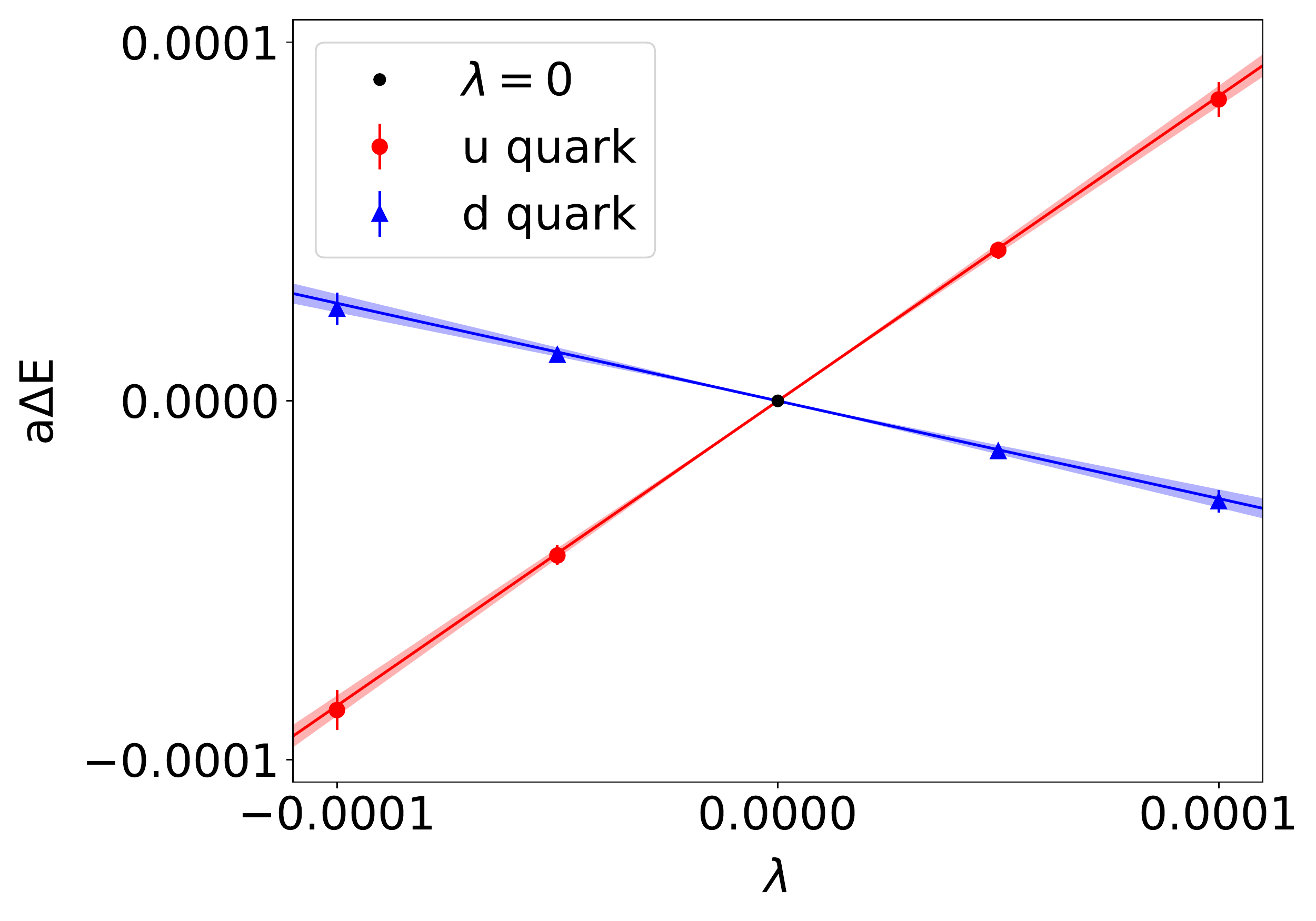}}
  \subfigure[]{\includegraphics[scale=0.23]{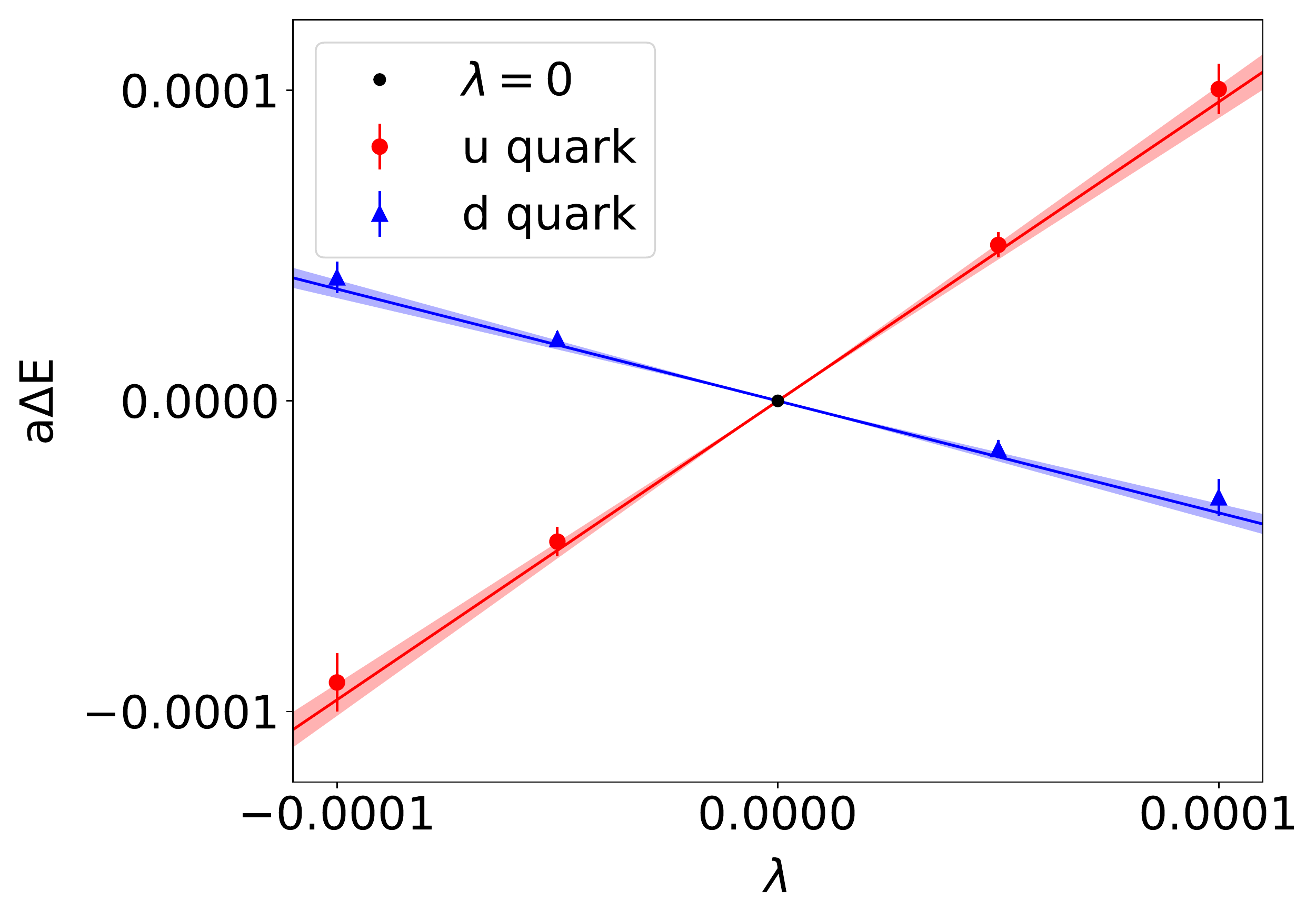}}
    \subfigure[]{\includegraphics[scale=0.23]{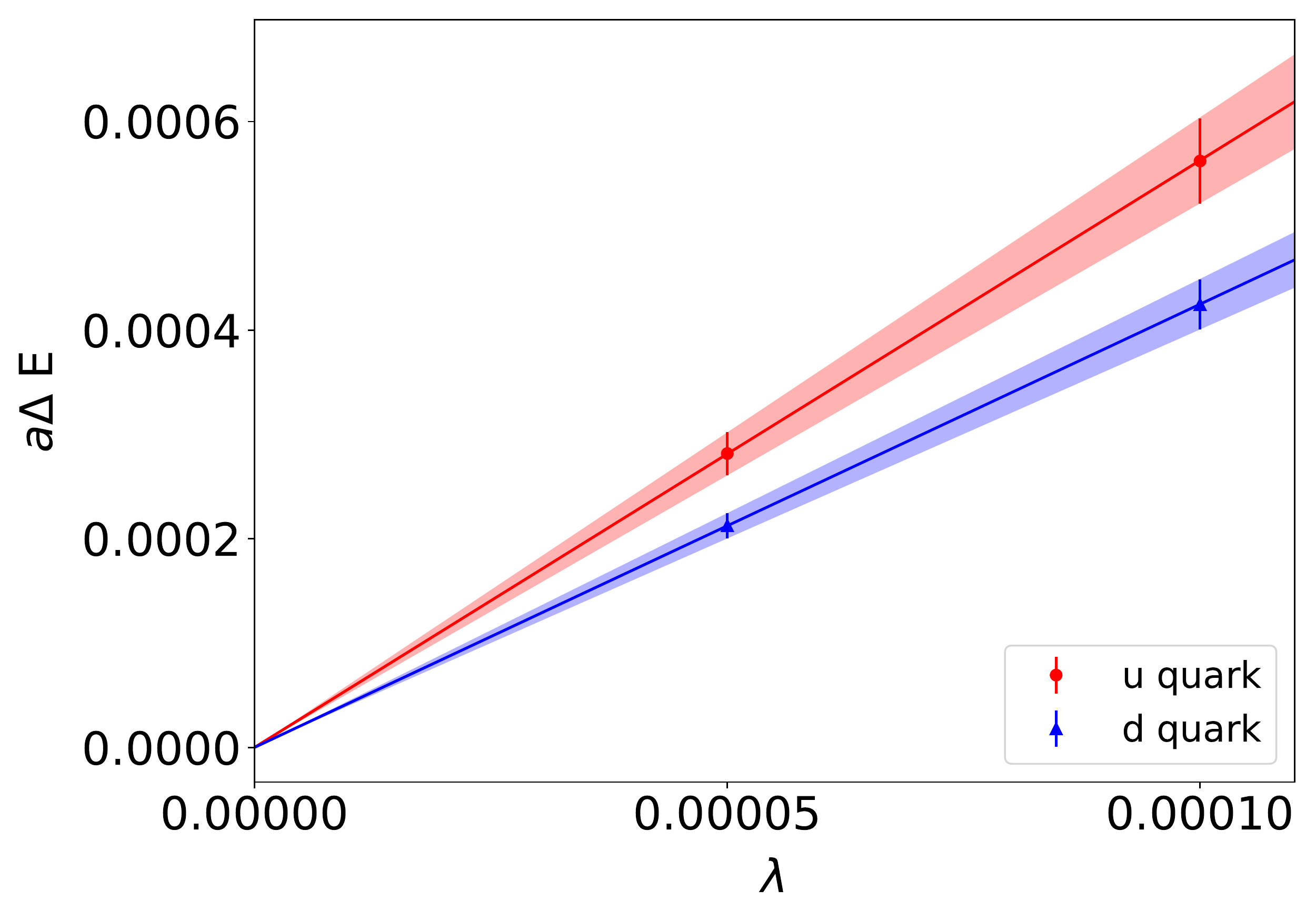}}
\caption{\label{fig:FH1} Proton energy shift, $\Delta E=E_\lambda-E$, for different parameter values, with a linear fit, where the red and blue bands show the statistical errors associated with the fitted parameters. Calculated at $a = 0.068$fm, $(\kappa_l,~\kappa_s)=(0.122167,~0.121682)$. Results for the tensor (a), axial (b) and scalar (c) operators}
\end{figure*}
\section{Weighted Averaging Method \label{sec:WA}}
The dependency of the fits on the time ranges used is a source of systematic uncertainty. To address these issues, we use a weighted averaging method on the fit results to limit the impact of the fit window selection. The weighted averaging method we use is a simplified variation of that outlined in detail in Ref.~\cite{PhysRevD.103.054504} and has similarities to that proposed in Ref.~\cite{PhysRevD.99.074510}. We proceed by determining the energy shifts, $\Delta E=E_\lambda- E$, by fitting the ratio of perturbed to unperturbed correlation functions using Eq.~\ref{eqn:corr_rat} for a variety of different Euclidean time fit windows. The largest time slice employed in each fit for each ensemble and operator is fixed to be the last time slice before the signal is lost due to statistical noise. For example, in Fig.~\ref{fig:FHeffmass} this would be chosen to be $t_{\text{max}}\approx17$. The start of the fit range, $t_{\text{min}}$, is varied between $t_{\text{min}/a}=6,~7,~8,~9,~10$ for ensembles with $\beta=5.40,~5.50,~5.65,~5.80,~5.95,$ respectively and up to the largest value of $t_{\text{min}}$ such that no less than four time slices are used in a fit. By adjusting the minimum time of the fit range, $t_{\text{min}}$, based on the lattice spacing of each ensemble, we are ensuring that each fit starts at an earlier scale. In the following we refer to the value of $\Delta E$ for a single fit, $f$, as $E^f$. Each fit result is then assigned a weight:
\begin{align}
    w^f=\frac{p_f(\delta E^f)^{-2}}{\sum^{N}_{f^\prime=1}p_{f^\prime}(\delta E^{f^\prime})^{-2}},
\end{align}
where $f$ labels the choice of fit range specified by $t_\text{min}$ for a fixed $t_\text{max}$, $p_f=\Gamma(N_{\text{dof}}/2,~\chi^2/2)/\Gamma(N_{\text{dof}}/2)$ is the p-value of the fit and $\delta E^{f}$ is the uncertainty in the energy shift, $E^f$, for fit $f$. Taking a weighted average of the $N$ fit findings, $E^f$, provides the final estimate of the energy shift, $\overline{E}$, and associated uncertainty $\delta \overline{E}$,
\begin{equation}
\begin{aligned}
    \overline{E}~=&~\sum^{N}_{f=1}w^f E^f,\\
    \delta_{\text{stat}}\overline{E}^2~=&~\sum^{N}_{f=1} w^f(\delta E^f)^2,\\
     \delta_{\text{sys}}\overline{E}^2~=&~\sum^{N}_{f=1} w^f(E^f-\overline{E})^2,\\
     \delta\overline{E}~=&~\sqrt{\delta_{\text{stat}}\overline{E}^2+\delta_{\text{sys}}\overline{E}^2}.
\end{aligned}
\end{equation}
The total error $\delta \overline{E}$ describes the combined statistical uncertainty on $\overline{E}$ plus the systematic uncertainty
arising from the choice of fit range. The separating of this error into $\delta_\text{stat}\overline{E}$ and $\delta_{\text{sys}}\overline{E}$ only partially separates statistical and systematic uncertainties because $\delta_\text{stat}\overline{E}$ includes statistical errors
plus systematic uncertainties related to fluctuations among the $\delta E^f$. The final estimate, $\overline{E}$, provides an estimate of the energy of the hadron with reduced systematic bias arising from choice of fit window. Fig.~\ref{fig:WARES} shows the proton effective energy shift for the ratio (Eq.~\ref{eqn:corr_rat}), using the standard definition of an effective mass. The final estimate of the energy shift, $\overline{E}$, when using the weighted averaging method is indicated by the red band. Fig.~\ref{fig:WARES} also includes a bar graph for the weights assigned to each fit value.
\section{Determination of Matrix Elements}
\subsection{Feynman-Hellman Method}
Now that we have a procedure for reliably determining the energy shifts, we are now in a position to determine $\Delta E$ at multiple values of $\lambda$ for a fixed ensemble and operator. In Fig.~\ref{fig:FH1} we plot the calculated proton energy shifts $\Delta E$ for each value of $\lambda$ for the $a = 0.068$fm ensemble with $(\kappa_l,~\kappa_s)=(0.122167,~0.121682)$. Fig.~\ref{fig:FH1}(a) shows results for the tensor operator, while Fig.~\ref{fig:FH1}(b) shows those for the axial operator. Now performing a linear fit to Eq.~\ref{eqn:DELTAE} and extracting the slope we obtain the following results,
$g^u_T=0.822(27)$, $g^d_T=-0.263(25)$, $g^u_A=0.814(56)$ and $g^d_A=-0.316(26)$, with the tensor results, renormalised at $\mu=2~\text{GeV}$ in the $\overline{\text{MS}}$ scheme using the renormalisation factors given in Table \ref{tab:renorm}.
Similarly for the scalar charge, in Fig.~\ref{fig:FH1}(c) we perform a linear fit to Eq.~\ref{eqn:DELTAEscalar} and by extracting the slope we find, $g^u_S=4.03(29)$ and $g^d_S=3.04(17)$, again renormalised at $\mu=2~\text{GeV}$ in the $\overline{\text{MS}}$ scheme. The above process has been repeated for all quark masses on each of the lattice spacings as well as for the $\Sigma$ and $\Xi$ baryons. The results can been found in Appendix \ref{appendix4}, in Tables \ref{tab:tensor}, \ref{tab:axial}, \ref{tab:scalar}.  \\
\subsection{Two-exponential fit method}
Here we compare the FH method results to the popular “two-exponential fit” method using three point functions. This is undertaken by expanding the two-point and three-point functions to the second energy state and fitting to obtain the parameters of interest. The process for the two-exponential fit is to fit the two-point correlator over a sink time range in which the two-state initial fit assumption is justified. Then using these extracted parameters in the fit to the three-point correlator using a $\tau$ range that also satisfies a two-state initial fit assumption. A detailed treatment of the two-exponential fit is given in, for example, Ref.~\cite{Dragos:2016rtx}. 
\begin{figure}[H]
\includegraphics[scale=0.45]{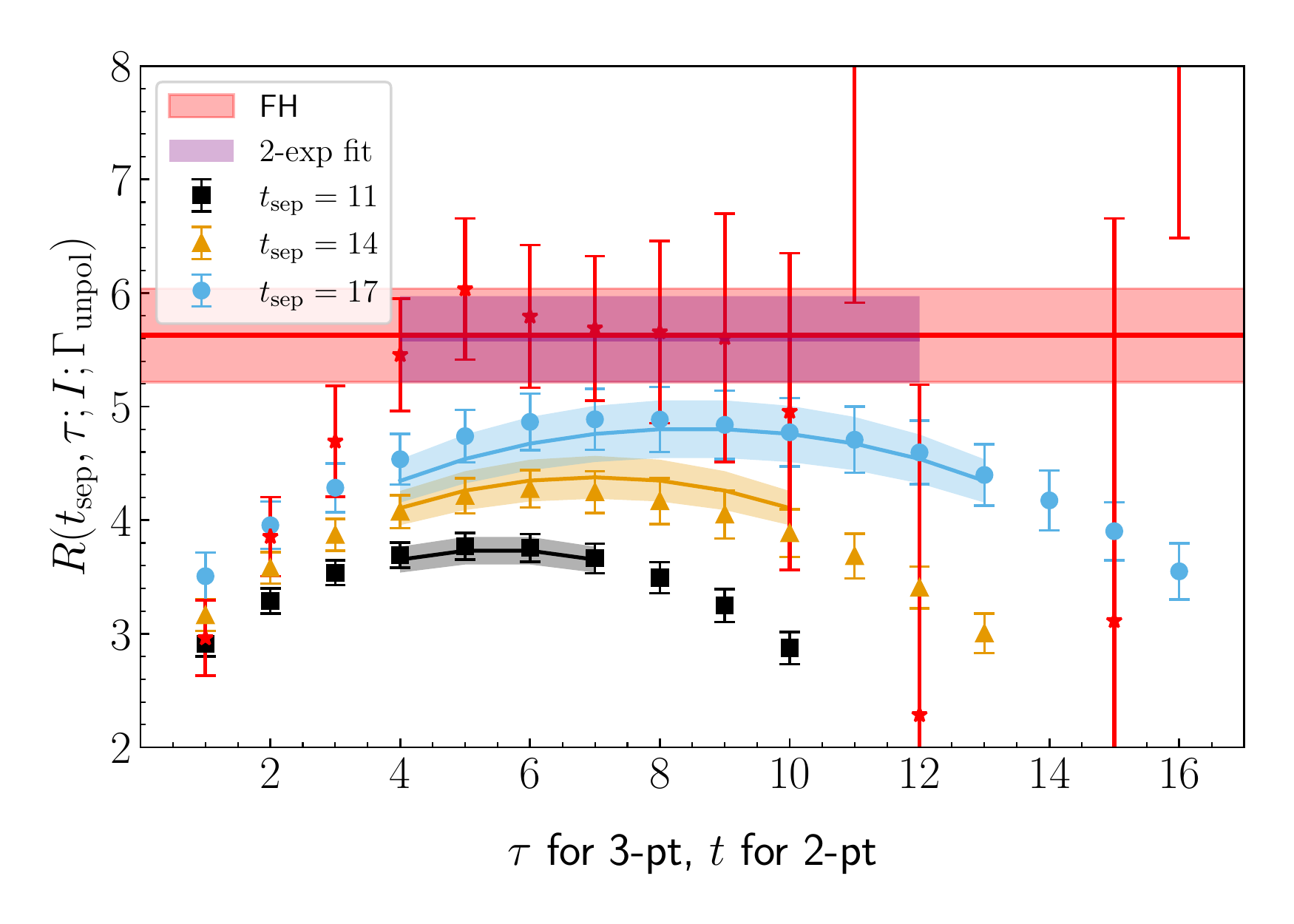}
\caption{\label{fig:exp2fit}Graph of $g_S^u$ extracted using the FH method shown by the red points and shaded band compared with the result using the two-exponential fit method, calculated at $a=0.068$fm, $(\kappa_l,~\kappa_s)=(0.122167,~0.121682)$. The black, orange and blue fits correspond to the two-exponential fit function constructed and the purple shaded area corresponds to the $g_S^u$ parameter extracted from the two-exponential fit. }
\end{figure}
Fig. \ref{fig:exp2fit} shows a comparison of the result for $g_S^u$ extracted using the FH method (red band) and the result using the two-exponential fit method (purple band). The red points come from a fixed $\lambda$ value, similar to that shown in Fig.~\ref{fig:FHeffmass}, whereas the red band comes from performing a linear fit to Eq.~\ref{eqn:DELTAE} and extracting the slope. We can see that the results using the FH method is in excellent agreement with the standard three-point analysis.\\ 

Now that we have the quark contributions for multiple lattice ensembles, in the next section we shall use a $SU(3)$ flavour symmetry breaking method to extrapolate results for the nucleon isovector charges  to the physical quark mass.
\section{Flavour Symmetry Breaking\label{sec:FB}}
The QCD interaction is flavour-blind, which means that the only distinction between quark flavours comes from the quark masses when we disregard the electromagnetic and weak interactions. The theory behind these interactions is easiest to understand when all three quark flavours share the same mass, as this allows us to use the full power of $SU(3)$ flavour symmetry. Here we have kept the bare quark mass, $\bar{m}=(m_u+m_d+m_s)/3$, held fixed at its physical value, while systematically varying the individual quark masses around the $SU(3)$ flavour symmetric point, $m_u=m_d=m_s$, in order to constrain the extrapolation to the physical point. In this work we simulate with degenerate $u$ and $d$ quark masses $m_u=m_d\equiv m_l$, restricting ourselves to $n_f=2+1$.\\

When $SU(3)$ is unbroken all octet baryon matrix elements of a given octet operator can be expressed in terms of just two couplings $f$ and $d$. However, once $SU(3)$ is broken and we move away from the symmetric point we can construct quantities ($D_i$, $F_i$) which are equal at the symmetric point but differ in the case where the quark masses are different. The theory behind constructing these quantities is described in detail in Ref.~\cite{Bickerton_2019} and is summarised below. The result of constructing these quantities leads to `fan' plots, with slope parameters ($r_i$, $s_i$) relating them. Following the method in Ref.~\cite{Bickerton_2019} we use the $SU(3)$ expansion to extrapolate the nucleon charges to the physical point.\\

In this work, we describe the quark mass dependence of the hadronic matrix elements by a perturbation in the quark masses about an $SU(3)$ symmetric point. This perturbation generates a polynomial expansion in the quark mass differences (i.e. the $SU(3)$ breaking parameter) and therefore appears distinct from a chiral formulation that generates nonanalytic behaviour (e.g. logarithms) in the vicinity of the 2- or 3-flavour chiral limits. However, it has been demonstrated in Ref.~\cite{PhysRevD.84.054509}, that by expanding the logarithmic features about a fixed quark mass point (such as the chosen $SU(3)$ symmetric point), the infrared singularities reveal themselves in the high-order terms of the polynomial expansion — hence demonstrating that the group-theoretic expansion does encode the same physics that appears in the logarithms. A detailed numerical investigation exploring the numerical convergence from both limits goes beyond the present work. Here we assess the convergence of our expansion empirically, subject to the precision of our numerical results.
\subsection{Mass dependence of amplitudes}
In order to find the allowed mass dependence of the octet operators in hadrons we need the $SU(3)$ decomposition of the $8\otimes8\otimes8$. $SU(3)$ singlet and octet coefficients are constructed through group theory and using a mass Taylor expansion, which can be seen in Ref.~\cite{Bickerton_2019}. Here we summarise the coefficients in Table \ref{tab:coeff}.\\
\begin{ruledtabular}
\begin{table}[ht]
\centering
\begin{tabular}{llll|lllll}
\multicolumn{4}{l}{} $1$,$1^{st}$ class& \multicolumn{5}{l}{} $8$, $1^{st}$ class   \\
\multicolumn{4}{l}{}  $\mathcal{O}(1)$ & \multicolumn{5}{l}{} $\mathcal{O}(\delta m_l)$ \\
 &  & \textit{f} &\textit{d} & \textit{d} &\textit{d}  &  \textit{d}& \textit{f} & \textit{f}  \\ 
\hline
 I& $A_{\bar{B}^\prime FB}$ &\textit{f}  &\textit{d}& $r_1$& $r_2$ & $r_3$ & $s_1$ & $s_2$  \\ 
\hline
 0& $\bar{N}\eta N$ & $\sqrt{3}$  &-1 & 1 & 0 & 0 & 0 &-1   \\
 0& $\bar{\Sigma}\eta\Sigma$& 0 &  2&1  & 0 & 2$\sqrt{3}$ & 0 & 0  \\
 0& $\bar{\Lambda}\eta\Lambda$ &0   &-2 & 1 & 2 & 0 & 0 &  0 \\
 0& $\bar{\Xi}\eta\Xi$ & -$\sqrt{3}$  &-1  &  1& 0 & 0 & 0 &   1\\ 
\hline
 1& $\bar{N}\pi^0 N$&  1&  $\sqrt{3}$ & 0 & 0 & -2 & 2 & 0  \\
 1& $\bar{\Sigma}\pi^0\Sigma$&2  & 0 &0  & 0 & 0 & -2 & $\sqrt{3}$  \\
 1& $\bar{\Xi}\pi^0\Xi$ &1 &-$\sqrt{3}$ &0  &0  &2  & 2 &0  
\end{tabular}
\caption{\label{tab:coeff} Coefficients in the mass Taylor expansion of $A_{\bar{B'}FB}$  operator amplitudes: $SU(3)$ singlet and octet, for first class currents \cite{Bickerton_2019}.} 
\end{table}
\end{ruledtabular}

These coefficients are used to construct equations which are linear in $\delta m_l$, where:
\begin{align}
  \delta m_{l}=m_l-\bar{m},
  \label{eq:deltaml}
\end{align} 
is the difference of the light quark mass to the $SU(3)$ symmetric point. 
Using the definitions in Table~\ref{tab:currents}, we introduce the notation for the matrix element transition of $B\rightarrow B^\prime$ as follows:
\begin{align}
A_{\bar{B'}FB}=\bra{B'}J^F\ket{B},
\end{align}
\begin{ruledtabular}
\begin{table}
\centering
\begin{tabular}{c|ccc}
 Index & Baryon ($B$) & Meson ($F$) & Current ($J^F$) \\ \hline
    1   &          $n$            &         $K^0$              &       $\bar{d}\gamma s$                \\ 
    2   &          $p$             &         $K^+$              &       $\bar{u}\gamma s$                \\ 
    3   &          $\Sigma^-$    &         $\pi^-$              &     $\bar{d}\gamma u$                  \\ 
    4   &          $\Sigma^0$    &         $\pi^0$              &     $\frac{1}{\sqrt{2}}\left(\bar{u}\gamma u - \bar{d}\gamma d \right)$                  \\ 
    5   &          $\Lambda^0$   &         $\eta$              &      $\frac{1}{\sqrt{6}}\left(\bar{u}\gamma u + \bar{d}\gamma d - 2\bar{s}\gamma s \right)$                 \\ 
    6   &          $\Sigma^+$    &         $\pi^+$              &     $\bar{u}\gamma d$                  \\ 
    7   &          $\Xi^-$       &         $K^-$              &       $\bar{s}\gamma u$                \\ 
    8   &          $\Xi^0$       &         $\bar{K}^0$              & $\bar{s}\gamma d$                      \\ \hline
    0   &                       &          $\eta^\prime$            &   $\frac{1}{\sqrt{6}}\left(\bar{u}\gamma u + \bar{d}\gamma d + \bar{s}\gamma s \right)$ 
\end{tabular}
\caption{\label{tab:currents} The conventions for the generalised currents. We use the convention that current (i.e. operator)
numbered by $i$ has the same effect as absorbing a meson with the index $i$. Here $\gamma$ represents an arbitrary Dirac matrix~\cite{Bickerton_2019}.}
\end{table}
\end{ruledtabular}
where $J^F$ is the appropriate operator, or current, from Table \ref{tab:currents} and $F$ represents the flavour structure of the operator. From Table \ref{tab:coeff} we can now read off the expansions of the various matrix elements, where the $f$ and $d$ terms are independent of $\delta m_l$ and the coefficients $r_1$, $r_2$, $r_3$ and $s_1$, $s_2$ are the leading order $\delta m_l$ terms. For example if we look at the $\bar{\Sigma}\pi\Sigma$ term, we have to first order in $\delta m_l$:
 \begin{align}
 \bra{\Sigma^+}J^{\pi^0}\ket{\Sigma^+}=A_{\bar{\Sigma}\pi\Sigma}=2f+(-2s_1+\sqrt{3}s_2)\delta m_l.
 \end{align} 
 \vspace{-10mm}
\subsection{Mass Dependence: `Fan Plots'}
\vspace{-3mm}
Since we hold the average quark mass, $\bar{m}$, fixed, while moving away from the symmetric point, we only need to consider the non-singlet polynomials in the quark mass. In this sub-section quantities $(D_i, F_i)$ are constructed which are equal at the symmetric point and differ in the case where the quark masses are different. We can then evaluate the the violation of $SU(3)$ symmetry that emerges from the difference in $m_s-m_l$. 
\subsubsection{The \textit{d}-fan} 
Following Ref.~\cite{Bickerton_2019}, we construct the following combinations of matrix elements which have the same value, $2d$, at the $SU(3)_d$ symmetric point:
\begin{equation}
\begin{aligned}
D_1\equiv-(A_{\bar{N}\eta N}+A_{\bar{\Xi}\eta \Xi})~=~&2d-r_1\delta m_l,\\
D_2\equiv A_{\bar{\Sigma}\eta \Sigma}~=~&2d+(r_1+2\sqrt{3}r_3)\delta m_l,\\
D_3\equiv -A_{\bar{\Lambda}\eta\Lambda}~=~&2d-(r_1+2r_2)\delta m_l,\\
D_4\equiv\frac{1}{\sqrt{3}}(A_{\bar{N}\pi N}-A_{\bar{\Xi}\pi \Xi})~=~&2d-\frac{4}{\sqrt{3}}r_3\delta m_l,\\
D_5\equiv A_{\bar{\Sigma}\pi\Lambda}~=~&2d+(r_2-\sqrt{3}r_3)\delta m_l,\\
D_6\equiv \frac{1}{\sqrt{6}}(A_{\bar{N}K\Sigma}+A_{\bar{\Sigma}K\Xi})~=~&2d+\frac{2}{\sqrt{3}}r_3\delta m_l,\\
D_7\equiv -(A_{\bar{N}K\Lambda}+A_{\bar{\Lambda}K\Xi})~=~&2d-2r_2\delta m_l.\\
\label{eqn:Dfan}
\end{aligned}
\end{equation}
By constructing these quantities the result is a `fan' plot with seven lines and three slope parameters $(r_1,r_2$ and $r_3)$ constraining them. The slope parameters can be constrained by calculating octet baryon matrix elements on a set of ensembles with varying quark masses at fixed lattice spacing, such as those given in Table~\ref{tab:para}, and constructing the $D_i$s. For the forward matrix elements considered here, these $D_i$s can also be written as linear combinations of the different quark contributions to the baryon charges.
For example, using Table~\ref{tab:currents} we see:
\begin{equation}
\begin{aligned}
D_1~=~&-(A_{\bar{N}\eta N}+A_{\bar{\Xi}\eta \Xi})\\
  =~&-\left(\frac{1}{\sqrt{6}}(g^u_p+g^d_p)+\frac{1}{\sqrt{6}}(g ^u_\Xi-2g^s_\Xi)\right),
\end{aligned}
\end{equation}
where we introduce the notation $g^q_B$ to denote the quark, $q$, contribution to the overall charge in the baryon, $B$. In this work we only consider the flavour diagonal matrix terms, i.e. there are no transition terms. Therefore, only the diagonal $D$ terms, $D_1$, $D_2$ and $D_4$, are used. An `average D' can also be constructed from the diagonal amplitudes:   
\begin{align}
X_D=\frac{1}{6}(D_1+2D_2+3D_4)=2d+\mathcal{O}(\delta m_l^2),
\label{eqn:avD}
\end{align}
 which is constant in $\delta m_l$ up to terms $\mathcal{O}(\delta m_l^2)$. When constructing these fan plots it is useful to plot $\tilde{D}_i=D_i/X_D$ to find the average fit to reduce statistical fluctuations.
\subsubsection{The \textit{f}-fan}
Similarly another five quantities, $F_i$, can be constructed which all have the same value, $2f$, at the $SU(3)_f$ symmetric point:
\begin{equation}
\begin{aligned}
F_1\equiv \frac{1}{\sqrt{3}}(A_{\bar{N}\eta N}-A_{\bar{\Xi}\eta \Xi})~=~&2f-\frac{2}{\sqrt{3}}s_2\delta m_l,\\
F_2\equiv (A_{\bar{N}\pi N}+A_{\bar{\Xi}\pi \Xi})~=~&2f+4s_1\delta m_l,\\
F_3\equiv A_{\bar{\Sigma}\pi\Sigma}~=~&2f+(-2s_1+\sqrt{3}s_2)\delta m_l,\\
F_4\equiv\frac{1}{\sqrt{2}}(A_{\bar{\Sigma}K \Xi}-A_{\bar{N}K \Sigma})~=~&2f-2s_1\delta m_l,\\
F_5\equiv \frac{1}{\sqrt{3}}(A_{\bar{\Lambda}K \Xi}-A_{\bar{N}K \Lambda})~=~&2f+\frac{2}{\sqrt{3}}(\sqrt{3}s_1-s_2)\delta m_l.\\
\label{eqn:Ffan}
\end{aligned}
\end{equation}
Again, an `average F' can be calculated through:  
\begin{align}
X_F=\frac{1}{6}(3F_1+F_2+2F_3)=2f+\mathcal{O}(\delta m_l^2).
\label{eqn:avF}
\end{align}
In this work, only the connected quark-line terms are computed. Quark-line disconnected terms only show up in the $r_1$ coefficient and $r_1^{\text{discon}}$ cancels in the case $g^{u-d}_{T,A,S}=g_{T,A,S}^u-g_{T,A,S}^d$. Unlike the $d$-fan, the $f$-fan to linear order, has no error from dropping the quark-line disconnected contributions, as none of the $r_i$ parameters appear in the $f$-fan.
\subsection{Fan Plot Results}
\vspace{-3mm}
Here we present results using the $a=0.068$fm ensemble. Results from other lattice spacings are similar. In Section~\ref{sec:GF}, we will extend this method to include all ensembles and present the final results for $g_{T,A,S}$.
\begin{figure}[H]
    \centering
    \includegraphics[scale=0.3]{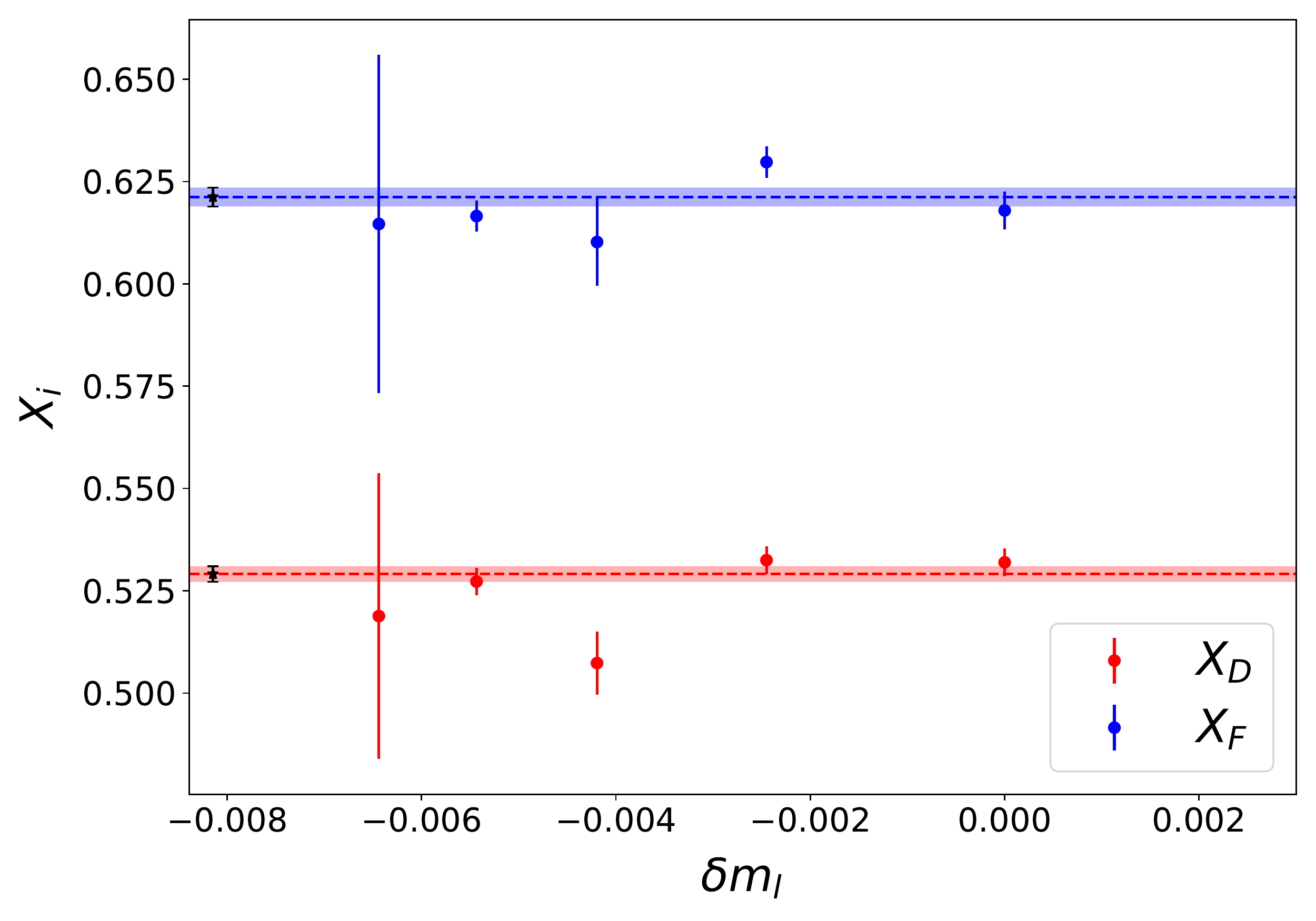}
    \caption{$X_D$ and $X_F$ for each $\delta m_l$ for the  $a=0.068$fm ensemble for the tensor matrix element. The dashed lines are constant fits and the black stars represent the physical point.}
    \label{fig:XDXF}
\end{figure}
\vspace{-5mm}
\begin{figure}[H]
  \centering
  \subfigure[]{\includegraphics[scale=0.27]{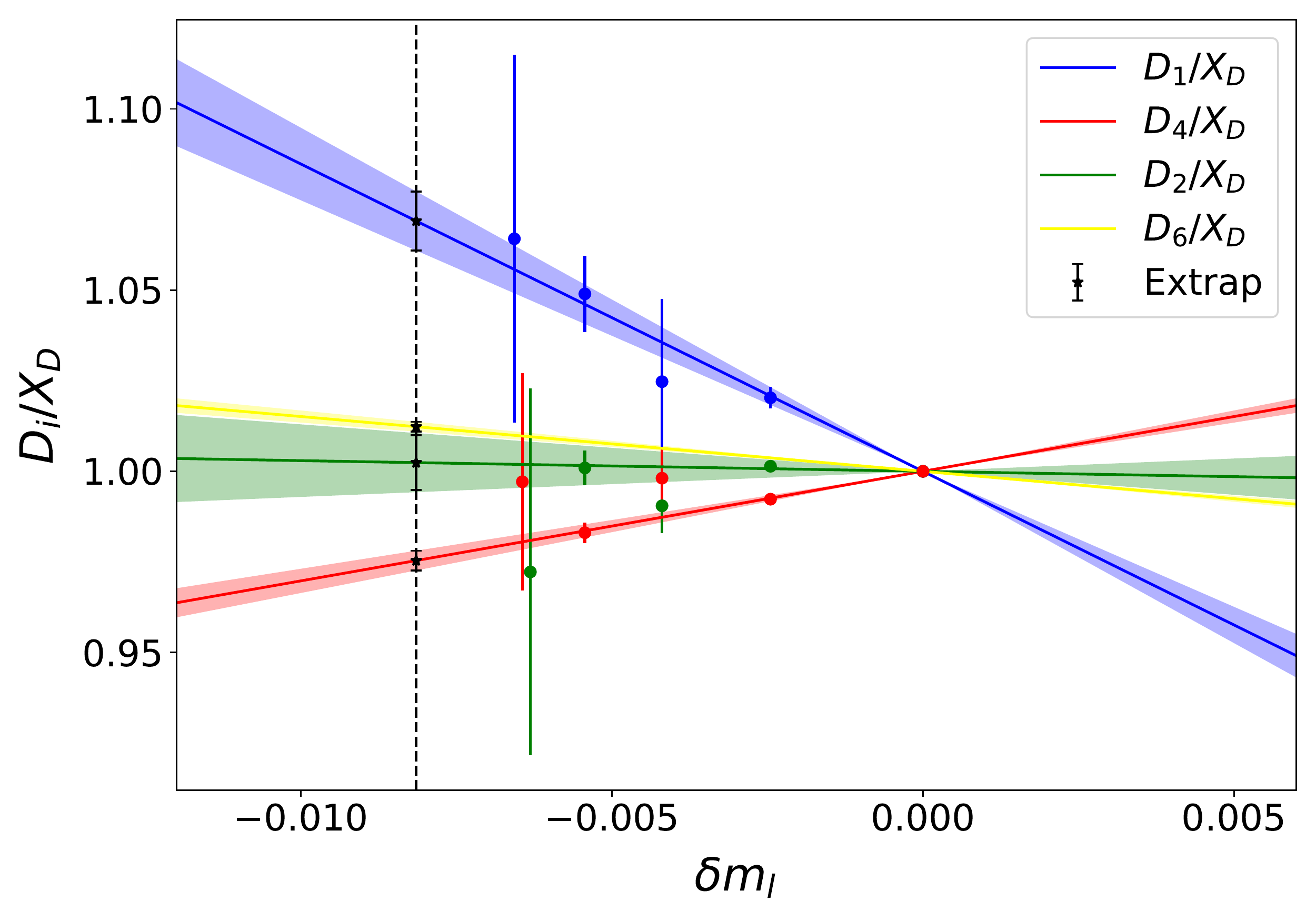}}
   \hskip -1ex
   \vspace{-3mm}
  \subfigure[]{\includegraphics[scale=0.27]{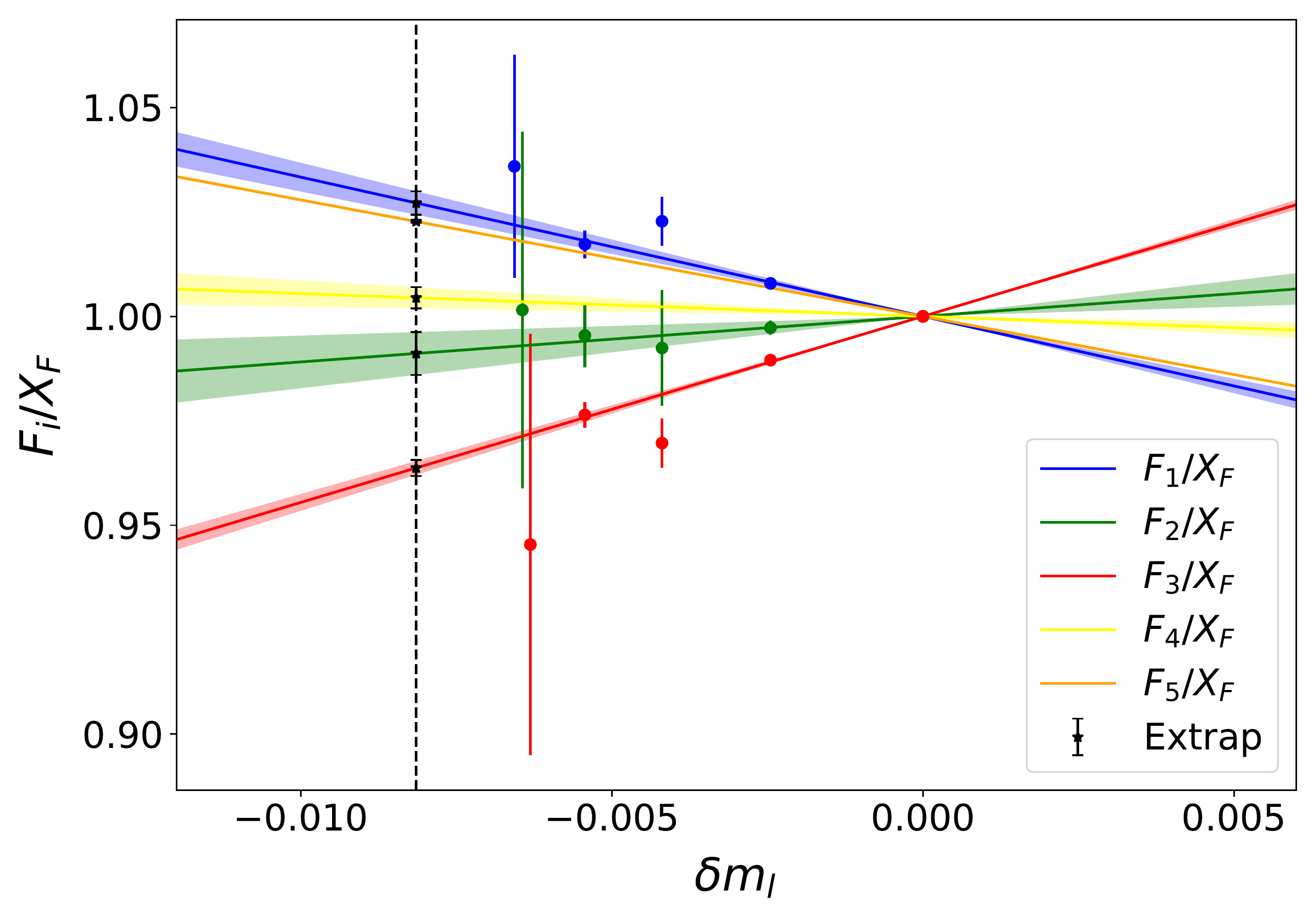}}
  \caption{\label{fig:fan}(a) The three fits $D_1$, $D_2$ and $D_4$ (b) The three fits $F_1$, $F_2$ and $F_3$ for the tensor. The vertical black dotted line represents the physical point. Results for the five ensembles at $a=0.068$fm ensemble. The flavour off-diagonal terms $D_6$, $F_4$ and $F_5$ are also predicted and plotted. Where some points have been offset slightly for clarity.}
\end{figure}
The singlet quantities $X_D$ and $X_F$ are calculated using Eq.~\ref{eqn:avD} and Eq.~\ref{eqn:avF}. In Fig.~\ref{fig:XDXF} $X_D$ and $X_F$ are plotted against $\delta m_l$ and fitted to a constant. Since in Section \ref{sec:GF} we will work to $\mathcal{O}(\delta m_l^2)$ in our flavour-breaking expansions, we fit $X_D$ and $X_F$ to constants in order to determine their values at the physical quark masses. The constant fits to the $a=0.068$fm data are shown by the dashed lines in Fig.~\ref{fig:XDXF}. In Fig.~\ref{fig:fan}(a) we present the D-`fan' plot which shows the $\delta m_l$ dependence of the $\tilde{D}_i=D_i/X_D$ for $i=1,~2$ and $4$. Here the lines correspond to the linear in $\delta m_l$ fits using Eq.~\ref{eqn:Dfan}. From these linear fits the slope parameters $\tilde{r}_1=r_1/X_D$ and $\tilde{r}_3=r_3/X_D$ are determined. It is interesting to note that these parameters also lead to a prediction for the flavour off-diagonal term for $i=6$, which is also shown. Similarly in Fig.~\ref{fig:fan}(b) we present the F-`fan' plot for $\tilde{F}_i=F_i/X_F$, $i=1,~2$ and $3$, where the lines correspond to the linear fits using Eq.~\ref{eqn:Ffan}. Similarly, the parameters $\tilde{s}_1=s_1/X_F$ and $\tilde{s}_2=s_2/X_F$ are determined from the linear fits. Again, the corresponding off-diagonal terms for $i=4,5$ are also predicted and plotted.
By forming appropriate linear combinations, we reconstruct the matrix elements for an individual quark flavour in a particular hadron: 
\begin{equation}
\begin{aligned}
\bra{p}\bar{u}\Gamma u\ket{p}~=&~2\sqrt{2}f+\Big(\sqrt{\frac{3}{2}}r_1-\sqrt{2}r_3+\sqrt{2}s_1\\&-\sqrt{\frac{3}{2}}s_2\Big)\delta m_l,&\\
  \bra{p}\bar{d}\Gamma d\ket{p}~=&~\sqrt{2}(f-\sqrt{3}d)+\Big(\sqrt{\frac{3}{2}}r_1+\sqrt{2}r_3-\sqrt{2}s_1\\&-\sqrt{\frac{3}{2}}s_2\Big)\delta m_l,&
\end{aligned}
\label{eq:u-d}
\end{equation}
and hence the nucleon isovector charges can be determined: 
\begin{equation}
\begin{aligned}
g^{u-d}_{T,A,S}=\bra{p}\bar{u}\Gamma u\ket{p}-\bra{p}\bar{d}\Gamma d\ket{p},
\label{eqn:final}
\end{aligned}
\end{equation}
for $\Gamma=\sigma_{34}\gamma_5,~\gamma_3\gamma_5$ and $I$. To obtain an extrapolation of $g_{T,A,S}$ to the physical point, we evaluate the expressions in Eq.~\ref{eq:u-d} at $\delta m_l \rightarrow\delta m_l^*$ and substitute in the estimated values for $r_i=\tilde{r}_i X_D$ and $s_i=\tilde{s}_i X_F$. In order to quantify systemic uncertainties we will now extend this flavour breaking expansion method further.
\section{Global Fits\label{sec:GF}}
The flavour breaking expansion described in Section \ref{sec:FB} only accounts for the quark mass-dependence of the matrix elements. However, in order to quantify systematic uncertainties, here we extend this method to also account for lattice spacing, finite volume effects and second order mass terms. As we are performing a global fit over all ensembles, we are now able to place constraints on the second-order mass terms, which means that all fits will now incorporate a term of order $\mathcal{O}(\delta m_l^2)$. These fits also include corrections with respect to $a$, $a^2$ and $m_\pi L$. In order to perform a global fit across all masses we substitute the quantity $\delta m_{l}$ from here on with:
\begin{align}
    \delta m_l \rightarrow \delta m_{l}=\frac{m_\pi^2-X_\pi^2}{X_\pi^2},
\end{align}
where the pseudoscalar mass flavour singlet, $X_\pi^2$, is given by:
\begin{align}
   X^2_\pi=\frac{2m_K^2+m_\pi^2}{3}.
\end{align}
By determining $\delta m_l$ to now be dimensionless and given in terms of physical quantities we are now able to combine results from different lattice spacings. The fit used for the singlet quantities $X_{D}$ and $X_{F}$ are extended to \cite{PhysRevD.84.054509}:
\begin{equation}
    \begin{aligned}
X_{D,F}=&X_{D,F}^*(1+c_1\frac{1}{3}[f_L(m_\pi)+2f_L(m_K)])+c_2a\\&+c_3\delta m_l^2,
\label{eqn:XDXF_gf}
\end{aligned}
\end{equation}
where we also consider an alternative $\mathcal{O}(a^2)$ lattice spacing dependence by replacing $c_2 a$ with $c_2 a^2$. The $c_1$ term estimates the finite size effects, where the leading meson-loop contribution has the functional form \cite{Beane:2004rf}:
\begin{align}
    f_L(m)=&\left(\frac{m}{X_\pi}\right)^2\frac{e^{-mL}}{\sqrt{mL}}.
\end{align}
It is important to note that here finite size effects are only included in the singlet quantities $X_D$ and $X_F$ and not in the $D$ and $F$ fan plot fits as the finite size corrections to the flavour-breaking coefficients determined by fits to, e.g. $\tilde{D}_i=D_i/X_D$ are expected to be sub-dominant compared to those in the corresponding singlet quantities.  The fits used for the $D$ fan, $\tilde{D}_i=D_i/X_D$, are of the form:
\begin{equation}
\begin{aligned}
\tilde{D}_1~=~&1-2(\tilde{r}_{1}+\tilde{b}_1 a)\delta m_l+\tilde{d}_1\delta m_l^2,\\
\tilde{D}_2~=~&1+((\tilde{r}_{1}+\tilde{b}_{1} a)\\&+2\sqrt{3}(\tilde{r}_{3}+\tilde{b}_{3} a))\delta m_l+\tilde{d}_{2}\delta m_l^2,\\
\tilde{D}_4~=~&1-\frac{4}{\sqrt{3}}(\tilde{r}_{3}+\tilde{b}_{3} a)\delta m_l+\tilde{d}_{4}\delta m_l^2,\\
\label{eqn:Dfan_gf}
\end{aligned}
\end{equation}
and similarly for the $F$ fan, $\tilde{F}_i=F_i/X_F$:
\begin{equation}
\begin{aligned}
\tilde{F}_1~=~&1-\frac{2}{\sqrt{3}}(\tilde{s}_2+\tilde{e}_2 a)\delta m_l +\tilde{f}_1\delta m_l^2,\\
\tilde{F}_2~=~&1+4(\tilde{s}_1+\tilde{e}_1 a)\delta m_l+\tilde{f}_2\delta m_l^2\\
\tilde{F}_3~=~&1+(-2(\tilde{s}_1+\tilde{e}_1 a)\\&+\sqrt{3}(\tilde{s}_2+\tilde{e}_2 a))\delta m_l+\tilde{f}_3\delta m_l^2.\\
\label{eqn:Ffan_gf}
\end{aligned}
\end{equation}
\begin{ruledtabular}
\begin{table*}
\centering
\begin{tabular}{ll|ll|ll|lll} 
\hline
 Fit&& $X_D$ &$\chi^2/dof$ &$X_F$&$\chi^2/dof$ & $g_T$ & $\chi^2/dof$ D-Fan&$\chi^2/dof$ F-Fan\\ 
\hline
1. & $\delta m_l^2$ & $0.515(43)$ & $1.88$ & $0.6002(57)$ & $1.74$ & $1.035(13)$ & $1.27$ & $1.84$ \\

2. & $a,~\delta m_l^2$ & $0.5251(81)$ & $1.87$ & $0.610(10)$ & $1.74$ & $1.000(27)$ & $0.76$ & $1.24$ \\

3. & $a^2,~\delta m_l^2$ & $0.5211(59)$ & $1.86$ & $0.608(69)$ & $1.74$ & $1.016(18)$ & $0.72$ & $1.22$ \\

4. & $a,~\delta m_l^2,~m_\pi L$ & $0.5252(80)$ & $1.98$ & $0.611(10)$ & $1.84$ & $1.001(27)$ & $1.35$ & $1.97$ \\

5. & $a^2,~\delta m_l^2,~m_\pi L$ & $0.5212(59)$ & $1.97$ & $0.606(75)$ & $1.84$ & $1.017(18)$ & $0.74$ & $1.18$ \\

6. & $\delta m_l^2,~m_\pi L$ & $0.516(43)$ & $1.98$ & $0.6005(50)$ & $1.83$ & $1.034(13)$ & $0.78$ & $1.21$ \\

\hline
 Fit&& $X_D$ &$\chi^2/dof$ &$X_F$&$\chi^2/dof$ & $g_A$ & $\chi^2/dof$ D-Fan&$\chi^2/dof$ F-Fan\\ 
\hline

1. & $\delta m_l^2$ & $0.583(21)$ & $0.99$ & $0.648(22)$ & $0.81$ & $1.262(60)$ & $1.00$ & $1.74$ \\

2. & $a,~\delta m_l^2$ & $0.565(36)$ & $1.02$ & $0.656(39)$ & $0.85$ & $1.21(15)$ & $1.03$ & $1.80$ \\

3. & $a^2,~\delta m_l^2$ & $0.572(26)$ & $1.02$ & $0.651(28)$ & $0.85$ & $1.231(95)$ & $1.02$ & $1.80$ \\

4. & $a,~\delta m_l^2,~m_\pi L$ & $0.563(36)$ & $1.08$ & $0.654(39)$ & $0.90$ & $1.21(15)$ & $0.93$ & $1.64$ \\

5. & $a^2,~\delta m_l^2,~m_\pi L$ & $0.574(27)$ & $1.08$ & $0.653(30)$ & $0.90$ & $1.231(95)$ & $0.95$ & $1.73$ \\

6. & $\delta m_l^2,~m_\pi L$ & $0.584(22)$ & $1.04$ & $0.648(22)$ & $0.85$ & $1.262(60)$ & $0.95$ & $1.73$ \\

\hline
 Fit&& $X_D$ &$\chi^2/dof$ &$X_F$&$\chi^2/dof$ & $g_S$ & $\chi^2/dof$ D-Fan&$\chi^2/dof$ F-Fan\\ 
\hline
1. & $\delta m_l^2$ & $-0.610(53)$ & $1.03$ & $2.52(12)$ & $1.52$ & $1.07(20)$ & $1.29$ & $2.76$ \\

2. & $a,~\delta m_l^2$ & $-0.72(10)$ & $0.98$ & $2.58(15)$ & $1.57$ & $1.12(50)$ & $1.30$ & $2.87$ \\

3. & $a^2,~\delta m_l^2$ & $-0.654(67)$ & $0.99$ & $2.55(13)$ & $1.57$ & $1.09(31)$ & $1.30$ & $2.89$ \\

4. & $a,~\delta m_l^2,~m_\pi L$ & $-0.71(10)$ & $1.04$ & $2.59(17)$ & $1.67$ & $1.11(50)$ & $1.07$ & $2.85$ \\

5. & $a^2,~\delta m_l^2,~m_\pi L$ & $-0.655(66)$ & $1.05$ & $2.52(14)$ & $1.66$ & $1.10(31)$ & $1.07$ & $2.99$ \\

6. & $\delta m_l^2,~m_\pi L$ & $-0.608(54)$ & $1.09$ & $2.51(12)$ & $1.60$ & $1.06(19)$ & $1.07$ & $2.98$ \\
\hline
\end{tabular}
\caption{\label{tab:resultsGF}Table of results for each fit and the corresponding $\chi^2/dof$, renormalised, where appropriate, at $\mu=2~\text{GeV}$ in the $\overline{\text{MS}}$ scheme. The notation in the first column shows which corrections are included in Eq.~\ref{eqn:XDXF_gf},~\ref{eqn:Dfan_gf} and~\ref{eqn:Ffan_gf}. For example Fit 4 includes all corrections $a$, $\delta m_{l}^2$ and $m_\pi L$, while Fit 1 only includes an added $\delta m_{l}^2$ term, i.e. $c_1=c_2=b_i=e_i=0$.}
\end{table*}
\end{ruledtabular}
\begin{figure*}
  \centering
  \subfigure[]{\includegraphics[scale=0.25]{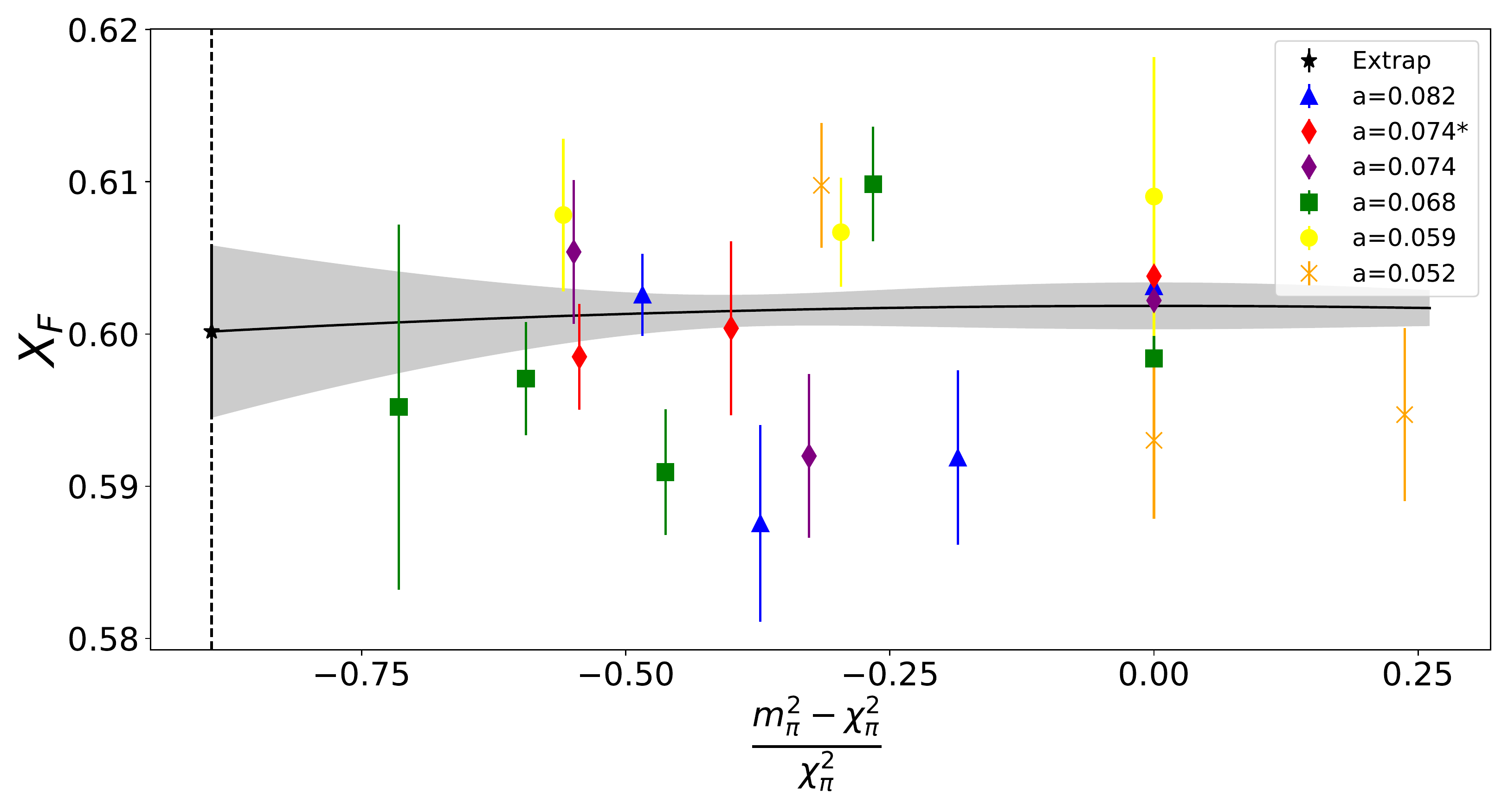}}
  \subfigure[]{\includegraphics[scale=0.25]{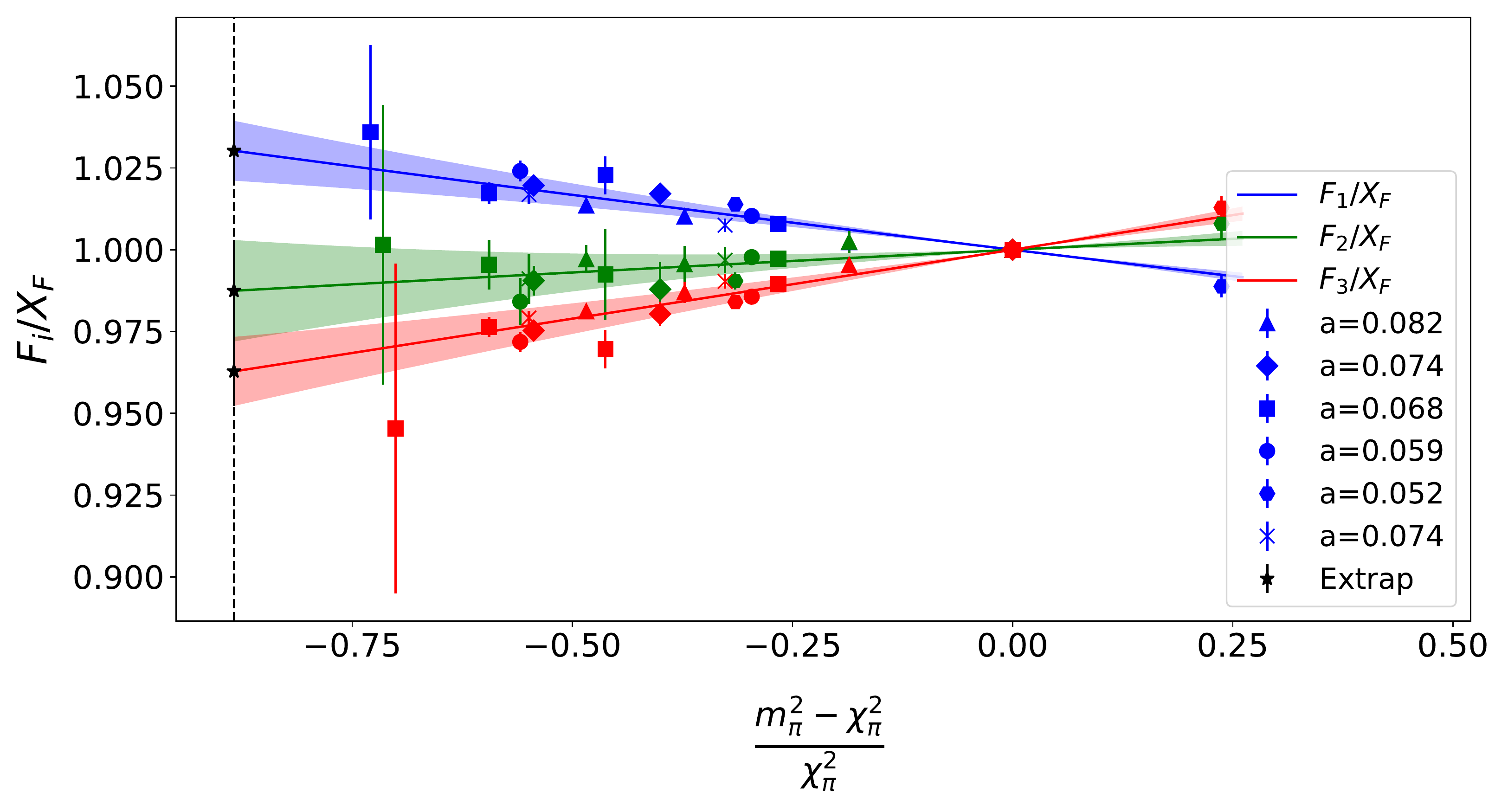}}
  \newline
    \subfigure[]{\includegraphics[scale=0.25]{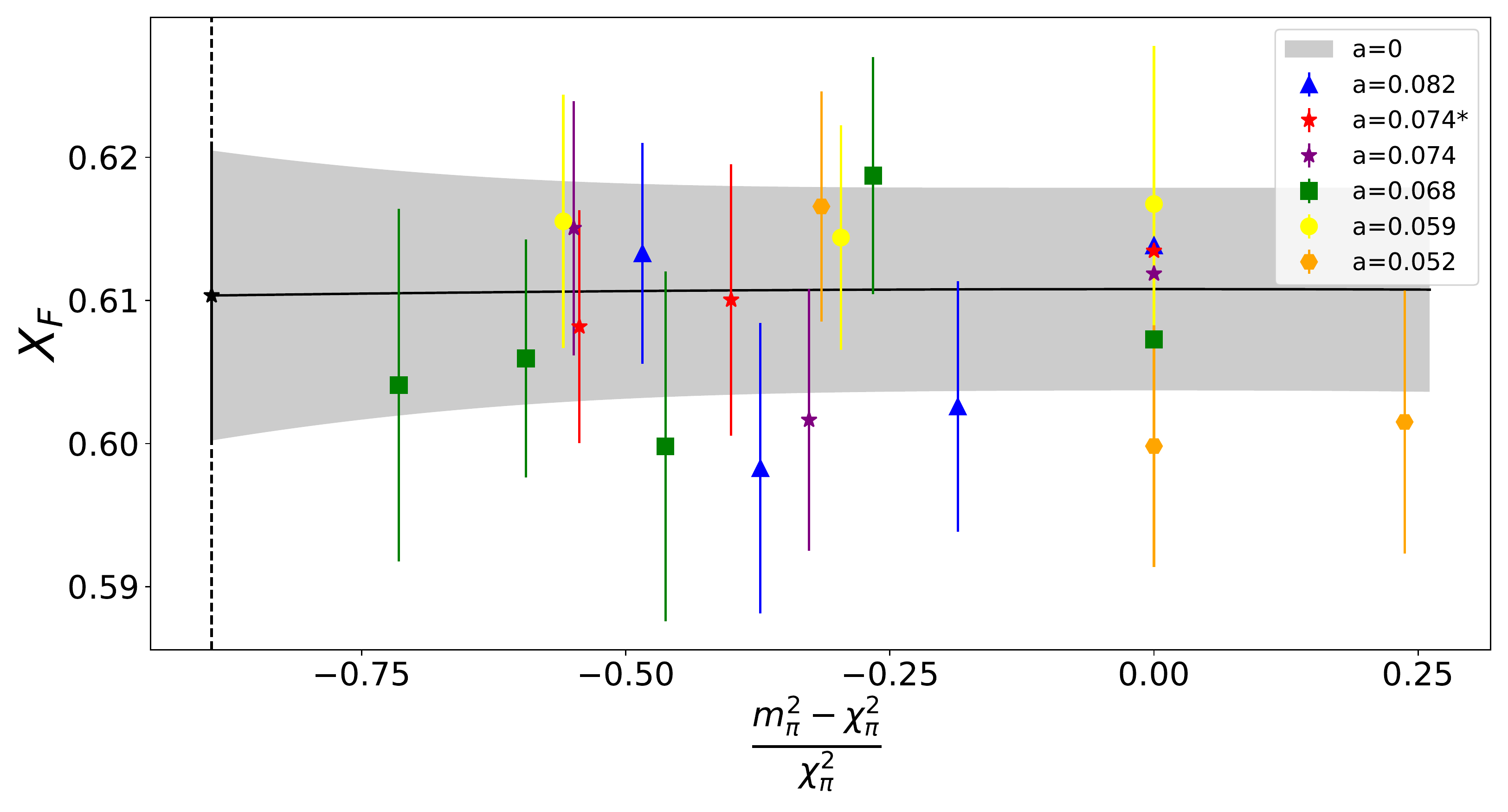}}
  \subfigure[]{\includegraphics[scale=0.25]{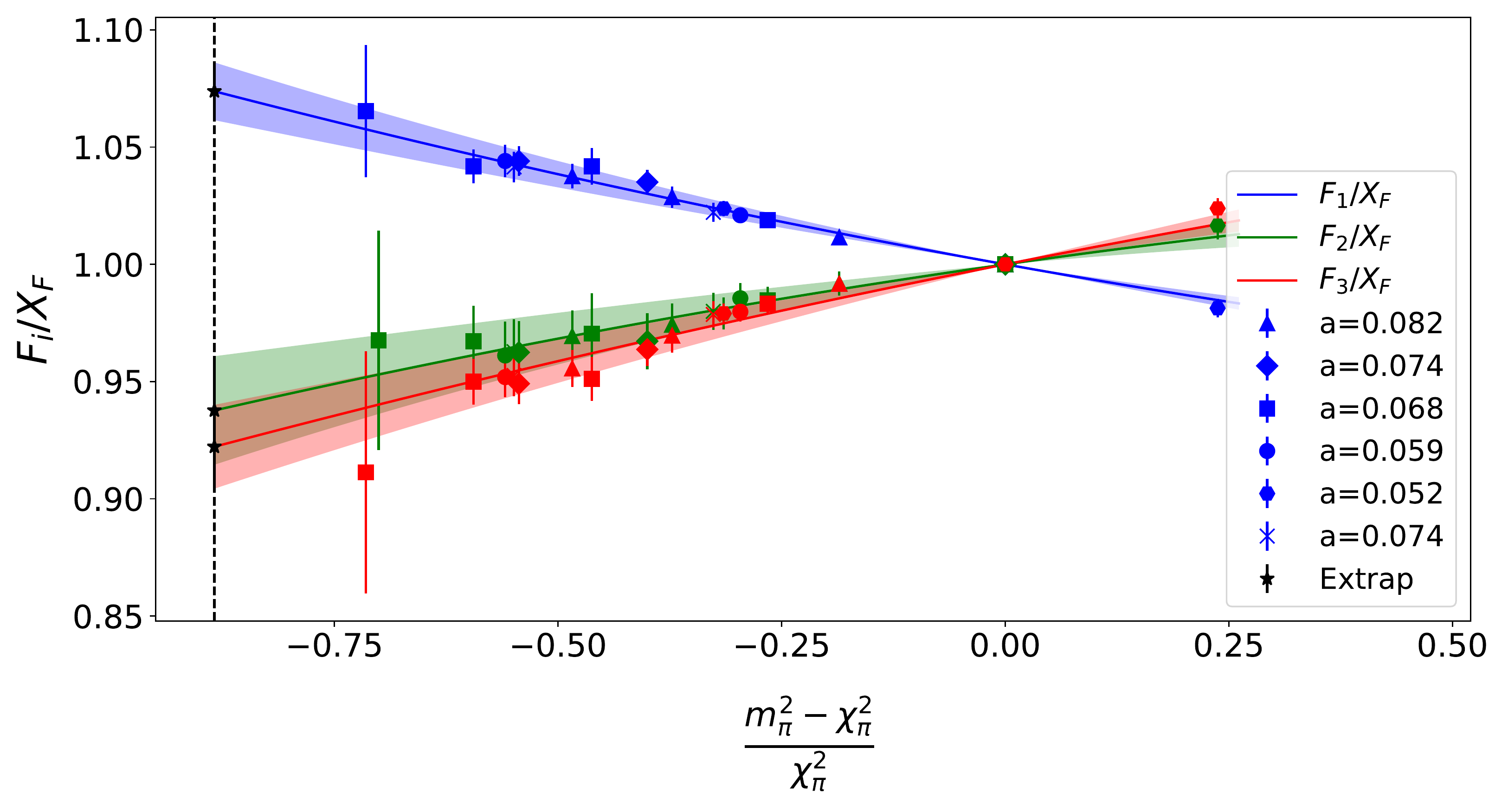}}
\caption{\label{fig:GFRESULTS}As an example of some fits we have for the tensor: (a) $X_F$ results for each ensemble using Eq.~\ref{eqn:XDXF_gf} where $c_1=c_2=0$ (Fit 1), plotted against $\frac{m_\pi^2-X_\pi^2}{X_\pi^2}$. (b) The three fits $F_1$, $F_2$ and $F_3$ using Eq.~\ref{eqn:Ffan_gf} with $e_i=0$ (Fit 1). (c) $X_F$ results using all corrections in Eq.~\ref{eqn:XDXF_gf} (Fit 4), plotted against $\frac{m_\pi^2-X_\pi^2}{X_\pi^2}$. The black line is a fit to Eq.~\ref{eqn:XDXF_gf} in the limit $a\rightarrow0$ and $m_\pi L\rightarrow \infty$. (d) The three fits $F_1$, $F_2$ and $F_3$ using Eq.~\ref{eqn:Ffan_gf}, where once again the data points are shifted in the limit $a\rightarrow0$. The black stars represent the physical point. Where some points have been offset slightly for clarity.}
\end{figure*}
\begin{figure*}
  \centering
  \subfigure[]{\includegraphics[scale=0.35]{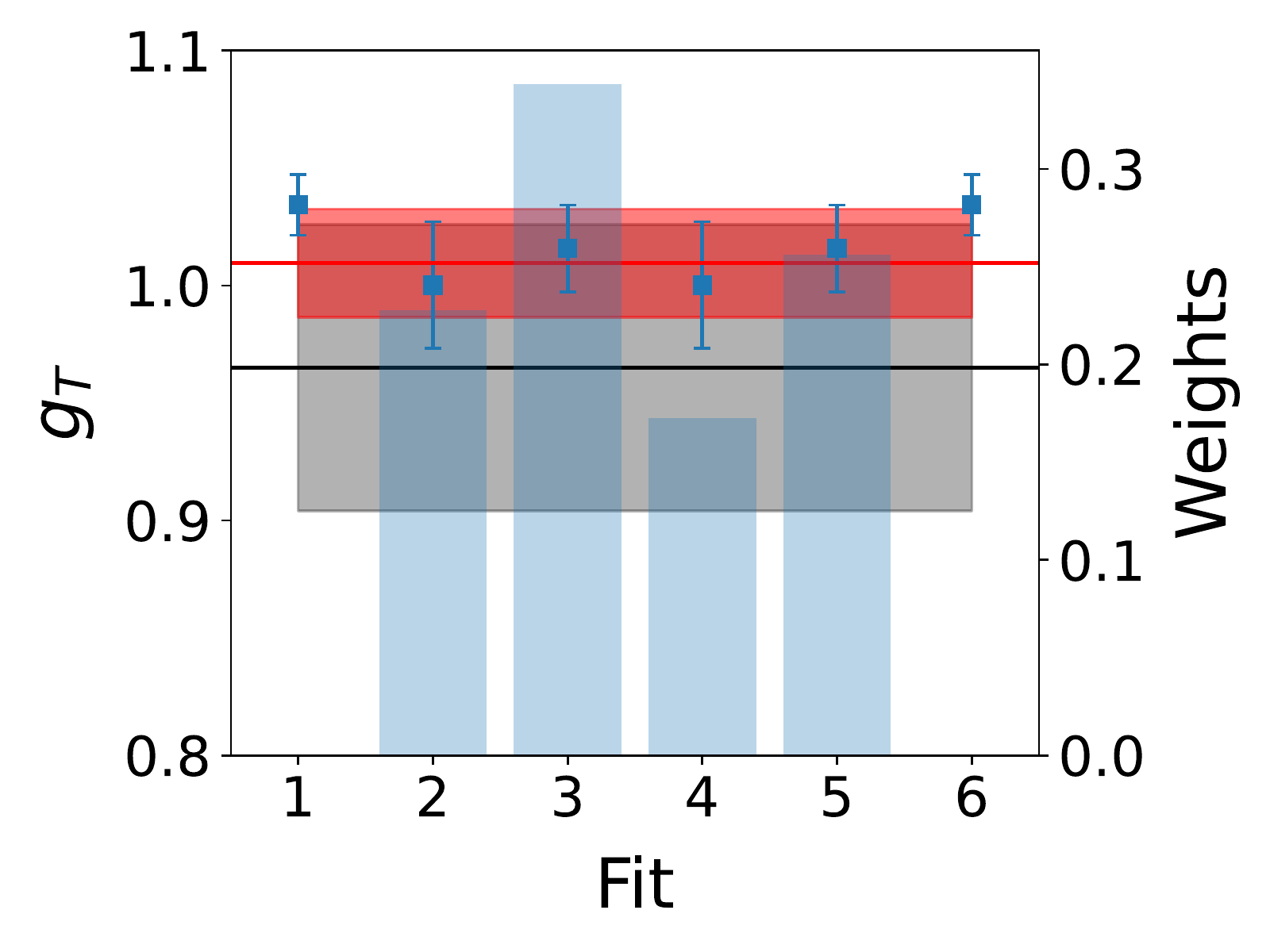}}
  \subfigure[]{\includegraphics[scale=0.35]{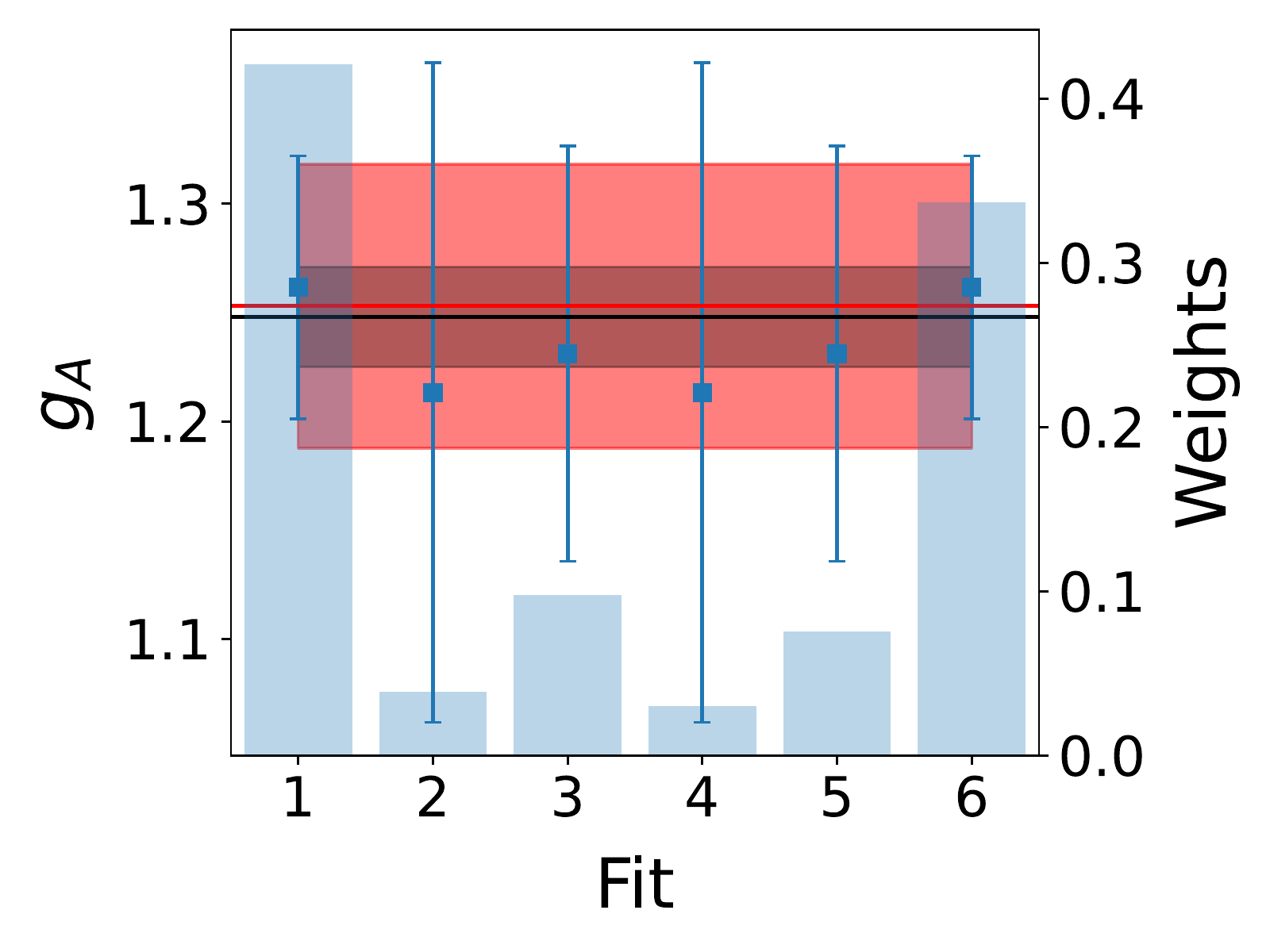}}
    \subfigure[]{\includegraphics[scale=0.35]{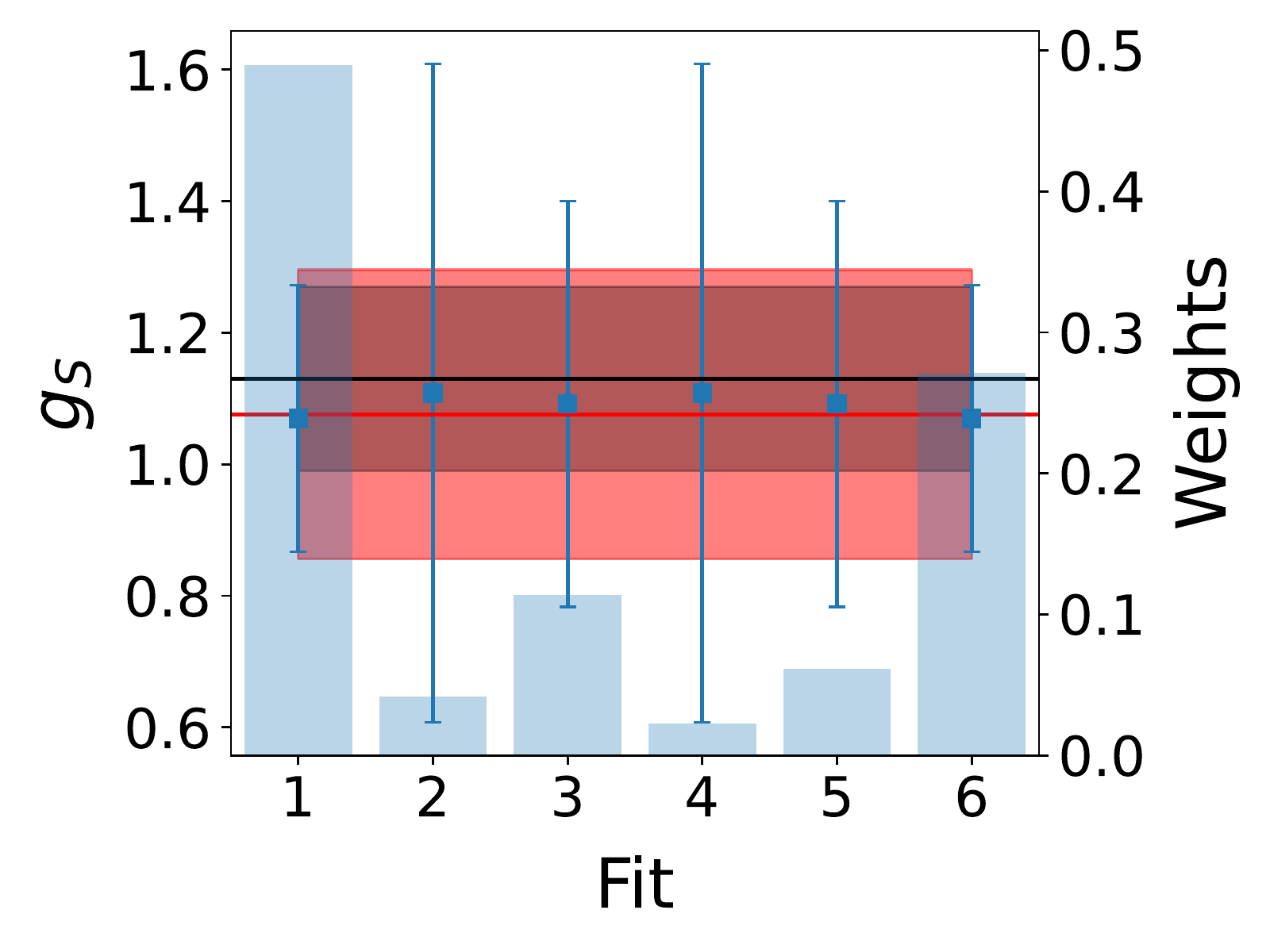}}
\caption{\label{fig:FINALRES} Weighted average results for $g_T$, $g_A$ and $g_S$. The $x$-axis displays the fit number as shown in Table~\ref{tab:resultsGF} and the $y$-axis displays the  corresponding nucleon isovector charge results. The bar graph shows the weight of each fit result. The red band shows the final weighted average result  using Eq. \ref{eqn:WAFIN}, with statistical and systematic errors combined in quadrature and the grey band is the FLAG Review result~\cite{FLAG2021}.}
\end{figure*}
The $\delta m_l^2$ coefficients were computed for the EM form factors in Ref.~\cite{Bickerton_2019}. At $\mathcal{O}(\delta m_l^2)$ there are 12 amplitudes and 11 coefficients so there is just one constraint. However, here we only consider the diagonal amplitudes and therefore we do not have 12 amplitudes and hence they are unable to be constrained here \cite{Horsley:2021zqe}. Therefore they are replaced with one $\delta m_l^2$ coefficient ($\tilde{d}_i,~\tilde{f}_i$) for each $D_i$ and $F_i$.\\ 

Now we perform a combination of different fits summarised in Table~\ref{tab:resultsGF}. 
Firstly, the fit is performed individually on $X_D$ and $X_F$. An example of this is shown in Fig.~\ref{fig:GFRESULTS}(a) and (c). In Fig~\ref{fig:GFRESULTS}(a) we show $X_F$ as a function $\frac{m_\pi^2-X_\pi^2}{X_\pi^2}$ for `Fit 1', which only includes the constant term, $X^*_F$ and a $\delta m_l^2$ term in Eq.~\ref{eqn:XDXF_gf}, while Fig~\ref{fig:GFRESULTS}(c) shows $X_F$ as a function of $\frac{m_\pi^2-X_\pi^2}{X_\pi^2}$ with the result from using Eq.~\ref{eqn:XDXF_gf} with all corrections included (`Fit 4'). The extrapolated result for $X_D$ and $X_F$ are summarised in Table~\ref{tab:resultsGF} taken in the limits $a\rightarrow 0$, $m_\pi L\rightarrow\infty$ and $m_\pi,~m_K \rightarrow$ physical masses. Similarly, fits are performed on the fan plots using Eq.~\ref{eqn:Dfan_gf} and Eq.~\ref{eqn:Ffan_gf}. Fig.~\ref{fig:GFRESULTS}(b) and (d) shows the results when using `Fit 1' and `Fit 4', where it is important to mention that all data points are shifted in the limit $a\rightarrow0$ in Figs.~\ref{fig:GFRESULTS}(c)(d). The slope results are then multiplied by the extrapolated results for $X_D$ and $X_F$:
\begin{equation}
   \begin{aligned}
    r_i=&(\tilde{r}_i+\tilde{b}_i a)X_D,\\
    s_i=&(\tilde{s}_i+\tilde{e}_i a)X_F,\\
    d_i=&\tilde{d}_i X_D,\\
    f_i=&\tilde{f}_i X_F.
\end{aligned} 
\end{equation}
The resulting slope parameters $r_i$, $s_i$ and the $\delta m_l^2$ coefficients are then included in the reconstruction of the matrix elements in a particular hadron:
\begin{equation*}
\begin{aligned}
\bra{p}\bar{u}\Gamma u\ket{p}~=&~2\sqrt{2}f+\Big(\sqrt{\frac{3}{2}}r_1-\sqrt{2}r_3+\sqrt{2}s_1\\&-\sqrt{\frac{3}{2}}s_2\Big)\delta m_l^*
+(-\frac{\sqrt{3}}{2\sqrt{2}}d_1+\frac{\sqrt{3}}{2\sqrt{2}}d_4\\&+ \frac{3}{2\sqrt{2}}f_1+\frac{1}{2\sqrt{2}}f_2)\delta m_l^{*2},
\end{aligned}
\end{equation*}
\begin{equation}
\begin{aligned}
  \bra{p}\bar{d}\Gamma d\ket{p}~=&~\sqrt{2}(f-\sqrt{3}d)+\Big(\sqrt{\frac{3}{2}}r_1+\sqrt{2}r_3-\sqrt{2}s_1\\&-\sqrt{\frac{3}{2}}s_2\Big)\delta m_l^* +(-\frac{\sqrt{3}}{2\sqrt{2}}d_1-\frac{\sqrt{3}}{2\sqrt{2}}d_4\\&+ \frac{3}{2\sqrt{2}}f_1-\frac{1}{2\sqrt{2}}f_2)\delta m_l^{*2},
  \label{eqn:endres}
\end{aligned}
\end{equation}
where $d=X_D^*/2$ and $f=X^*_F/2$. The final result for $g_{T,A,S}$ are then given in the limit, $a\rightarrow 0$, $m_\pi L\rightarrow\infty$ and $\delta m_l^*$ is the physical mass. The final results for $X_D$, $X_F$ and $g_{T,A,S}$ for each fit are summarised are in Table~\ref{tab:resultsGF}, together with the $\chi^2_{\text{reduced}}$ for each fit.
\subsection{Results}
In order to combine these results we extend our weighted averaging method described in section~\ref{sec:WA}. To do this we combine the $\chi^2$ and degrees of freedom of $X_D$, $X_F$, $D$-fan and $F$-fan; enumerated by $i=1,~2,~3,~4$, respectively, in the following: 
\begin{align}
\chi^2_f=\sum_{i=1}^{4}\chi^2_i,~~N_{\text{dof},f}=\sum_{i=1}^{4}N_{\text{dof},i},
\label{eqn:chi2}
\end{align}
where $f$ labels one of the six fit types. Each fit is then assigned a weight using the combined $\chi^2_f$:
\begin{align}
    \tilde{w}^f=\frac{p_f(\delta g_{T,A,S}^{f})^{-2}}{\sum^{6}_{f^\prime=1}p_{f^\prime}(\delta g_{T,A,S}^{f^\prime})^{-2}},
\end{align}
where $p_f=\Gamma(N_{\text{dof},f}/2,~\chi_f^2/2)/\Gamma(N_{\text{dof},f}/2)$ is the $p$-value of the fit $f$ and $\delta g_{T,A,S}^{f}$ is the uncertainty in the nucleon isovector charges calculated using Eq.~\ref{eqn:endres}. Taking a weighted average of the six fit results, $g^f_{T,A,S}$, provides a final estimate of the nucleon isovector charges, $g_{T,A,S}$, and associated uncertainty: 
\begin{equation}
\begin{aligned}
    \overline{g}_{T,A,S}~=&~\sum^{6}_{f=1}w^f g_{T,A,S}^f,\\
    \delta_{\text{stat}}\overline{g}_{T,A,S}^2~=&~\sum^{6}_{f=1} w^f(\delta g_{T,A,S}^f)^2,\\
     \delta_{\text{sys}}\overline{g}_{T,A,S}^2~=&~\sum^{6}_{f=1} w^f(g_{T,A,S}^f-\overline{g}_{T,A,S})^2,\\
     \delta\overline{g}_{T,A,S}~=&~\sqrt{\delta_{\text{stat}}\overline{g}_{T,A,S}^2+\delta_{\text{sys}}\overline{g}_{T,A,S}^2}.
     \label{eqn:WAFIN}
\end{aligned}
\end{equation}
Fig.~\ref{fig:FINALRES} shows the results for each fit and their assigned weight. The final estimate of the nucleon isovector charges, $\overline{g}_{T,A,S}$, renormalised using the results given in Table \ref{tab:renorm}, at $\mu=2~\text{GeV}$ in the $\overline{\text{MS}}$, are:
\begin{align}
    g_T~=&~1.010(21)_{\text{stat}}(12)_{\text{a}}(01)_{\text{FV}},\label{eqn:FINALresgt}\\
    g_A~=&~1.253(63)_{\text{stat}}(41)_{\text{a}}(03)_{\text{FV}},\label{eqn:FINALresga}\\
    g_S~=&~1.08(21)_{\text{stat}}(03)_{\text{a}}(00)_{\text{FV}},\label{eqn:FINALresgs}
\end{align}
where the systematic errors labelled as `a' and `FV' represent the difference in the central value obtained by incorporating a lattice spacing correction compared to without, and likewise for the finite volume correction. These final results, with statistical and systematic errors combined in quadrature, are shown by the red bands in Fig.~\ref{fig:FINALRES}. We note our results for $g_T$, $g_A$ and $g_S$ are all comparable with the FLAG Review results~\cite{FLAG2021}, represented by the grey bands in Fig.~\ref{fig:FINALRES}. Of particular note is that we have determined $g_T$ to the $\approx 2\%$ level. However, work is still needed in order reduce the uncertainties on, $g_S$ and $g_A$, to understand it at the same level.\\

As a check on our method for combining the results from the six different fits given in Table \ref{tab:resultsGF}, we employ the widely used Akaike Information Criterion (AIC). Here results obtained from the various fits are weighted using the Akaike weights~\cite{10.2307/2988185}:
\begin{align}
    w_f=\frac{\text{exp}(-\frac{1}{2}\text{AIC}_f(\Gamma))}{\sum_{f'}\text{exp}(-\frac{1}{2}\text{AIC}_{f'}(\Gamma))}.
\end{align}
Akaike's information criterion takes on the simple form for models with normally distributed errors:
\begin{align}
    \text{AIC}_f(\Gamma)=\chi^2_f+2p_f,
\end{align}
where $\chi^2_f$ is the same as that calculated in Eq.~\ref{eqn:chi2} and $p_f$ is the number of parameters in each fit. As a result the AIC weight prefers the models with lower $\chi^2$ values, but penalises those with too many fit parameters. The above method was repeated using the AIC weights. This gives the following results for the nucleon isovector charges, $g_T=1.003(26)$, $g_A=1.261(68)$ and $g_S=1.07(23)$, where the errors have been added in quadrature. These results are in agreement with those in Eq.~\ref{eqn:FINALresgt}, \ref{eqn:FINALresga} and \ref{eqn:FINALresgs}.
\subsection{Hyperons}
\vspace{-4mm}
Here we calculate flavour-diagonal matrix elements of hyperons using the same method. Ref.~\cite{Chang:2014iba} demonstrates that isovector combinations of hyperon charges are relevant in searches for new physics through semileptonic hyperon decays. The calculated slope parameters $r_i$, $s_i$ and the $\delta m_l^2$ coefficients can also be used in the reconstruction of the matrix elements in a particular hyperon. The theory behind constructing these quantities is described in detail in Ref.~\cite{Bickerton_2019} and is summarised in Appendix.~\ref{appendix1}. The results for the charges of the $\Sigma^+$ and $\Xi^0$ baryons are summarised in Table \ref{tab:hyper}.\\
\begin{ruledtabular}
\begin{table}[h]
\begin{tabular}{l|lll}
 &  Tensor& Axial & Scalar \\ \hline
 $g^u_{\Sigma^+}$& $0.802(16)(12)$ & $0.884(25)(36)$ & $2.75(25)(08)$ \\
 $g^s_{\Sigma^+}$& $-0.2379(10)(08)$ & $-0.250(22)(30)$ & $1.86(16)(12)$ \\\hline
 $g^u_{\Xi^0}$& $-0.1929(77)(13)$ & $-0.198(22)(15)$ & $1.52(11)(08)$ \\
 $g^s_{\Xi^0}$& $0.968(25)(10)$ & $0.924(23)(12)$ & $2.58(24)(11)$ \\
\end{tabular}
\caption{\label{tab:hyper} Summary of results for the tensor, axial and scalar charges of the $\Sigma^+$ and $\Xi^0$ baryons. The first set of brackets contains the statistical uncertainty, whereas the second set of brackets contains the systematic uncertainty.}
\end{table}
\end{ruledtabular}

To properly exploit the increased experimental sensitivity to hypothetical tensor and scalar interactions, we require lattice-QCD estimates of the nucleon isovector charge, $g_T$ at the level of $10–20\%$ \cite{Bhattacharya:2011qm}. The results presented here are at the $\delta g_T/g_T\approx 2\%$ level. As the overall goal of this research is to support precision tests of the Standard Model, we have successfully demonstrated the validity of our approach. We can now look at the effect this has on phenomenology.
\vspace{-5mm}
\section{Impact of Lattice Results on Phenomenology}
As discussed in Section~\ref{sec:intro}, it is expected that future neutron beta decay experiments will increase their sensitivity to BSM scalar and tensor interactions through improved measurements of the Fierz interference term, $b$, as well as the neutrino asymmetry parameter, $B$. In order to assess the full impact of these future experiments we have performed an analysis of the tensor charge $g_T$ and $g_S$. Here we discuss existing constraints on new scalar $\epsilon_S$ and tensor $\epsilon_T$ couplings which arise from low-energy experiments. Finally, using the existing constraints on $\epsilon_S$ and $\epsilon_T$ as well as our calculated value for $g_T$ and $g_S$, we determine the allowed regions in the $\epsilon_S-\epsilon_T$ plane.
\vspace{-5mm}
\subsection{Low-energy phenomenology of scalar and tensor interactions}
\subsubsection{$0^+\rightarrow0^+$ transitions and scalar interactions}
The most precise bound on the scalar coupling $\epsilon_S$ comes from $0^+\rightarrow0^+$ nuclear beta decay. The differential decay rate for $0^+\rightarrow0^+$ nuclear beta decay has coefficient $a_{0+}$ and Fierz interference term $b_{0+}$ \cite{Bhattacharya:2011qm}:  
\begin{align}
a_{0+}~=&~1,\\
b_{0+}~=&~-2\gamma g_S\epsilon_S,~~~~\gamma=\sqrt{1-\alpha^2Z^2},
\label{eqn:b0}
\end{align}
where $Z$ is the atomic number of the daughter nucleus. We can see from Eq.~\ref{eqn:b0} that $b_{0+}$ couples to the BSM scalar interaction. From a comparison of well known half-lives corrected by a phase-space factor, Hardy and Towner \cite{Hardy:2008gy} found $b_{0+}=-0.0022(26)$. This result was found using a number of daughter nuclei and averaging over the set. This can be converted to the following bound on the product of scalar charge and the new-physics effective scalar coupling:
\begin{align}
-1.0\times10^{-3}<g_S\epsilon_S<3.2\times10^{-3}~~~~(90\%~C.L.).
\label{eqn:dec}
\end{align}     
This is the most precise bound on the scalar interactions from low-energy probes.
\subsubsection{Radioactive Pion Decay and the Tensor Interaction}
\begin{figure}[H]
  \centering
  \includegraphics[scale=0.3]{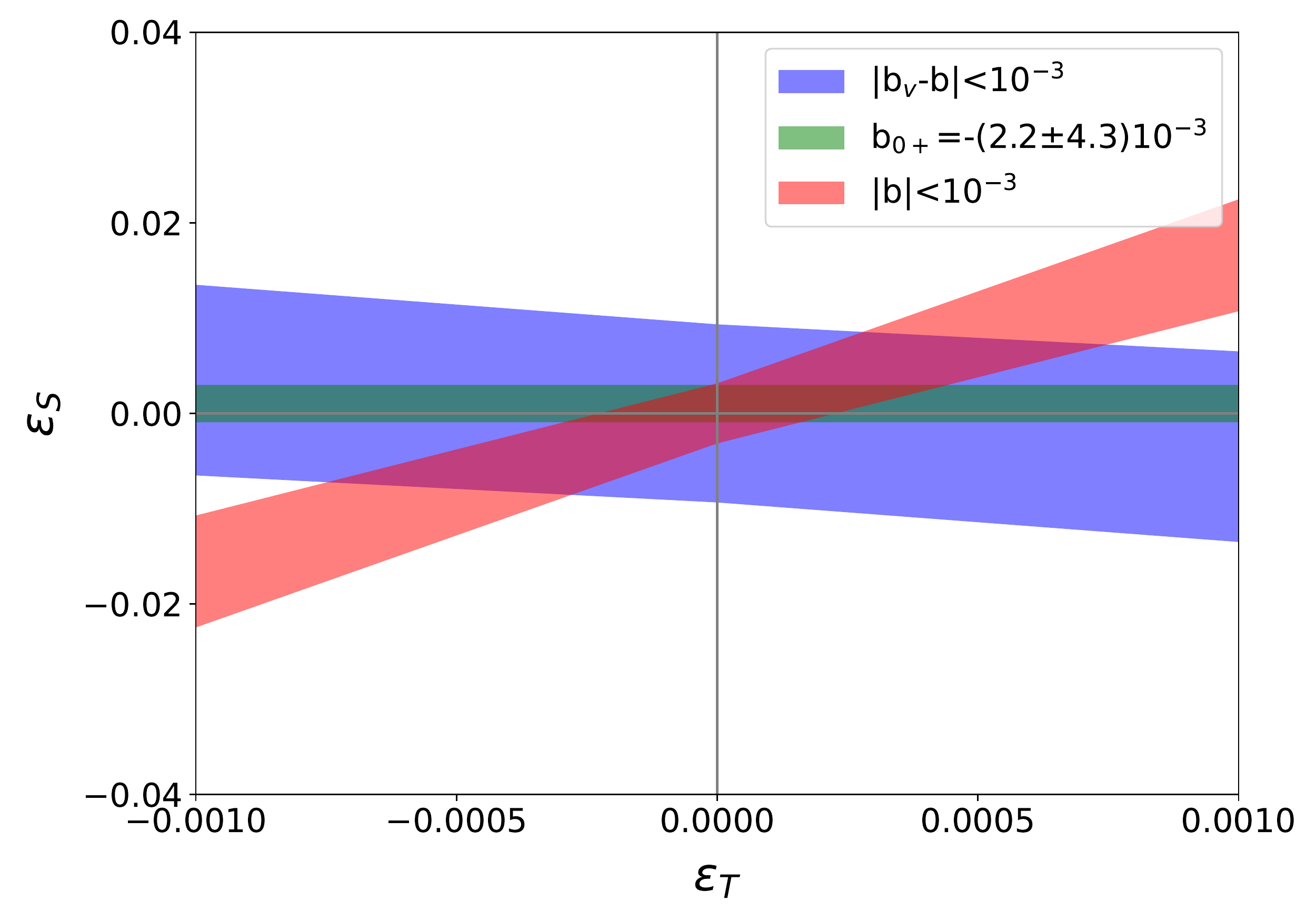}
  \caption{\label{fig:bound} Allowed regions in the $\epsilon_S-\epsilon_T$ plane, using the tensor and scalar charges as obtained in this work in Eq.~\ref{eqn:FINALresgt} and Eq.~\ref{eqn:FINALresgs}, $g_S=1.08(21)_{\text{stat}}(03)_{\text{sys}}$ and $g_T=1.010(21)_{\text{stat}}(12)_{\text{sys}}$. The green band is the existing band on $b_{0+}$ \cite{Bhattacharya:2011qm, Hardy:2008gy}.}
\end{figure}

An analysis of radioactive pion decay $\pi^+\rightarrow e^+\nu_e\gamma$ is sensitive to the same tensor operator that can be investigated in beta decays. The experimental results from the PIBETA collaboration \cite{Bychkov:2008ws} put constraints on $\epsilon_T$:
\begin{align}
-1.1\times10^{-3}<\epsilon_T<1.36\times10^{-3}~~~~(90\%~C.L.).
\label{eqn:pidec}
\end{align}  
Currently this is the most stringent constraint on the tensor coupling from low energy experiments. Using these constraints, as well Eq.~\ref{eqn:b} and Eq.~\ref{eqn:bv}, bounds can be put on the new scalar and tensor interactions at the $10^{-3}$ level.
Following the work of Ref.~\cite{Bhattacharya:2011qm}, in Fig.~\ref{fig:bound} we show the constraint on the $\epsilon_S-\epsilon_T$ plane.
The current best constraints on scalar and tensor interactions arise from $0^+\rightarrow0^+$ nuclear beta decays and radioactive pion decay, which is shown by the green band~\cite{Bhattacharya:2011qm, Hardy:2008gy}. The neutron constraints are future projections at the $10^{-3}$ level, derived from Eq. \ref{eqn:b} and Eq. \ref{eqn:bv}, using the tensor and scalar charges as obtained in this work, shown by the red and blue bands in Fig. \ref{fig:bound}. When accounting for uncertainties in these lattice QCD calculations, the boundaries on the bands in Fig.~\ref{fig:bound} become wider and the bands take on a `bow-tie' shape. However most of the constraining power is lost due to the large uncertainty in our value for $g_S$.
In order to fully utilise the constraining power of $10^{-3}$ experiments, understanding the lattice-QCD estimates of the nucleon tensor and scalar charge at the level of $10\%$ is required \cite{Bhattacharya:2011qm}. We have successfully calculated the tensor charge at the $\approx2\%$ level and are able to fully utilise the constraining power future experiments.
\subsection{Quark electric dipole moment}
In this section we briefly discuss the impact our results have on constraining the quark EDM couplings using the current bound on the neutron EDM. Using the same method followed in Section \ref{sec:GF} we are able to constrain $g_T^q$. 
We note that in this work we have only considered quark-line connected contributions, although other works have shown the disconnected contributions to be small at near-physical quark masses \cite{PhysRevD.98.091501}. This is in line with expectations based on the fact that the tensor operator is a helicity-flip operator and hence disconnected contributions mush vanish in the chiral limit. Using Eq.~\ref{eqn:endres} we can calculate the up and down contributions to the nucleon tensor charge for each fit listed in Table ~\ref{tab:resultsGF}. Applying the weighted averaging method, the final estimates for, $g^T_q$, are:
\begin{align}
    g_T^u~=~0.812(21),\\
    g_T^d~=~-0.199(14).
\end{align}
\begin{figure}[H]
\includegraphics[scale=0.3]{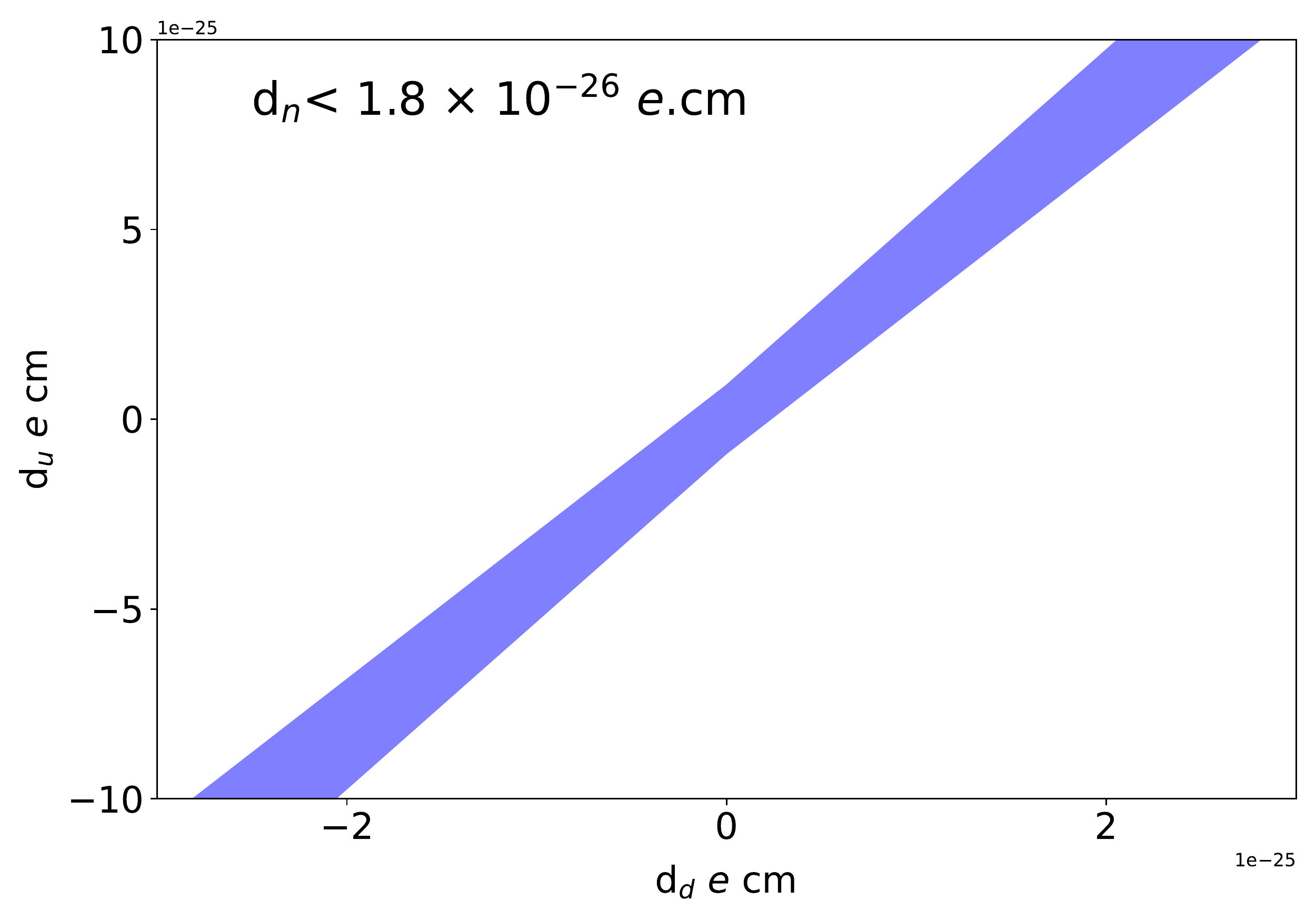}
\caption{\label{fig:dnbounds} $90\%$ confidence level bounds on $d_u$ and $d_d$ using lattice QCD estimates for $g_T^u$ and $g_T^d$ and the current limit on the neutron EDM of $|d_n| < 1.8\times10^{-26} e$.cm~\cite{PhysRevLett.124.081803}.}
\end{figure}
Using these results, Eq.~\ref{eqn:dn} and the existing bound on the neutron EDM we are able to put bounds on the new effective couplings which contain new CP violating interactions. Fig.~\ref{fig:dnbounds} shows the $90\%$ confidence level bounds in the $d_u-d_d$ plane, assuming $g_T^s=0$. 
\section{Conclusion}
In this work we have presented results for the axial, tensor and scalar nucleon and hyperon charges using the Feynman-Hellmann theorem, as well as using a flavour symmetry breaking method to systematically approach the physical quark masses. We applied a weighted averaging method on the fit results, removing possible systematic uncertainties which arise from a bias in choosing the fit windows. In the flavour symmetry breaking method, symmetry constraints are automatically built in order-by-order in $SU(3)$ breaking. We extended the flavour symmetry breaking method in this analysis in order to have full coverage of $a$, $m_\pi$ and volume, meaning we have control over these systematics. Our final result of $g_T=1.010(21)_{\text{stat}}(12)_{\text{sys}}$ is comparable to results present in the FLAG review. We have precisely calculated $g_T$ to the $\approx 2\%$ level, successfully reaching the goal of understanding $g_T$ at the $10\%$ level. However, work is still needed in order reduce the error on, $g_S = 1.08(21)_{\text{stat}}(03)_{\text{sys}}$ and $g_A=1.253(63)_{\text{stat}}(41)_{\text{sys}}$, to understand it at the same level. Future work is still needed with access to physical quark masses in order to better constrain the extrapolation to the physical point.
\begin{acknowledgments}
The numerical configuration generation (using the BQCD lattice QCD 
program \cite{Haar}) and data analysis (using the Chroma software 
library \cite{EDWARDS2005832}) was carried out on the 
DiRAC Blue Gene Q and Extreme Scaling Service (EPCC, Edinburgh, UK), 
the Data Intensive Service (Cambridge Service for Data-Driven Discovery,
CSD3, Cambridge, UK), the Gauss Centre for Supercomputing (GCS) 
supercomputers JUQUEEN and JUWELS (John von Neumann Institute for Computing,
NIC, J\"ulich, Germany) and resources provided by the 
North-German Supercomputer Alliance (HLRN),
the National Computer Infrastructure (NCI National Facility in Canberra, 
Australia supported by the Australian Commonwealth Government) and the 
Phoenix HPC service (University of Adelaide). 
R.H. is supported in part by the STFC grant ST/P000630/1.
P.E.L.R. is supported in part by the STFC grant ST/G00062X/1. 
G.S. is supported by DFG grant SCHI 179/8-1.
R.D.Y. and J.M.Z. are supported by the ARC grants DP190100298 and DP220103098.
\end{acknowledgments}

\appendix
\section{Lattice Ensemble Details}
\label{appendix2}
\begin{ruledtabular}
\begin{table}[H]
\centering
\begin{tabular}{lllll}
 $\beta$& $a$(fm) &Volume & $(\kappa_\text{light},\kappa_\text{strange})$  & $\#$Trajectories\\ 
\hline
 $5.40$& $0.082$ & $32^3 \times 64$ &$(~0.119930~,~0.119930~)$  & $1639$\\
 &  &  & $(~0.119989~,~0.119812~)$ & $1005~~(2)$ \\
 &  &  & $(~0.120048~,~0.119695~)$ & $1000~~(3)$\\
 &  &  & $(~0.120084~,~0.119623~)$ & $1345~~(3)$\\ 
\hline
 $5.50$& $0.074$ & $32^3 \times 64$ & $(~0.120900~,~0.120900~)$ & $1754$\\
 &  &  & $(~0.121040~,~0.120620~)$ & $1216$  \\
 &  &  & $(~0.121095~,~0.120512~)$&  $1849~~(2)$\\
\hline
 $5.50$& $0.074$ & $32^3 \times 64$ & $(~0.120950~,~0.120950~)$ & $1614$\\
 &  &  & $(~0.121040~,~0.120770~)$ & $1762$ \\
 &  &  & $(~0.121099~,~0.120653~)$&  $1003~~(2)$ \\
 \hline
 $5.65$& $0.068$ & $48^3\times 96$ & $(~0.122005~,~0.122005~)$ &  $531$ \\
 &  &  & $(~0.122078~,~0.121859~)$ &  $633$ \\
 &  &  &  $(~0.122130~,~0.121756~)$&  $561~~~(2)$ \\
 &  &  & $(~0.122167~,~0.121682~)$ & $534~~~(2)$  \\
 &  & $64^3\times 96$ & $(~0.122197~,~0.121623~)$ & $428~~~(3)$\\
\hline
 $5.80$& $0.059$ & $48^3\times 96$ & $(~0.122810~,~0.122810~)$ &  $298$\\
 &  &  & $(~0.122880~,~0.122670~)$ &  $458~~~(2)$ \\
 &  &  & $(~0.122940~,~0.122551~)$ & $522$\\ 
\hline
 $5.95$ & $0.052$& $48^3\times 96$ & $(~0.123411~,~0.123558~)$ & $283~~~(2)$ \\
 &  &  & $(~0.123460~,~0.123460~)$ & $457$ \\
 &  &  & $(~0.123523~,~0.123334~)$ & $415$
\end{tabular}
\caption{\label{tab:samples} Details of the lattice ensembles used in this work: the same number of configurations was used for each $\lambda$ value and operator. Measurements are separated by a single HMC trajectory with a randomised source location. The number in parentheses indicates the quantity of randomised sources used per configuration to generate additional samples.}
\end{table}
\end{ruledtabular}
\section{Individual quark contributions to the overall charge in the baryon.}
\label{appendix4}
Here we present the the bare results for the individual quark contributions to the overall tensor, axial and scarlar charges in the nucleon, $\Sigma$ and $\Xi$ baryons.
\begin{ruledtabular}
\begin{table*}[h!]
\centering
\begin{tabular}{ll|llllll} 
$\beta$ & $\kappa_l$  & $g_{T_P}^u$ & $g_{T_P}^d$ & $g_{T_\Sigma}^u$ & $g_{T_\Sigma}^s$ & $g_{T_\Xi}^u$ &  $g_{T_\Xi}^s$\\
\hline
 $5.40$&$0.119930$  & $0.8851(55)$ & $-0.2020(43)$  & $0.8851(55)$ & $-0.2020(43)$ &$-0.2020(43)$  &   $0.8851(55)$ \\
 & $0.119989$ &$0.832(23)$  & $-0.222(10)$ &$0.838(25)$  & $-0.216(18)$ & $-0.222(18)$ & $0.851(24)$  \\
 & $0.120048$ & $0.849(24)$ & $-0.225(17)$ & $0.845(25)$ & $-0.2145(83)$ & $-0.209(11)$ &  $0.870(11)$  \\
 & $0.120084$ & $0.842(32)$ & $-0.209(25)$ & $0.830(17)$ & $-0.2098(61)$ & $-0.2112(97)$ & $0.8760(83)$ \\
 \hline
 $5.50$& $0.120900$ & $0.869(10)$ & $-0.2145(34)$ & $0.869(10)$ & $-0.2145(34)$ & $-0.2145(34)$ & $0.869(10)$ \\
 & $0.121040$  &$0.810(38)$  &$-0.202(22)$ & $0.809(25)$ & $-0.2163(98)$ & $-0.2042(83)$ & $0.8796(85)$ \\
 & $0.121095$  & $0.800(27)$ &$-0.198(22)$  & $0.822(17)$ & $-0.2159(66)$ &$-0.1947(59)$  &  $0.8747(61)$\\
 \hline
$5.50$& $0.120950$  & $0.8830(59)$ & $-0.2115(39)$ & $0.8830(59)$  & $-0.2115(39)$ & $-0.2115(39)$ &$0.8830(59)$   \\
 & $0.121040$  & $0.863(11)$ & $-0.2066(46)$  & $0.8597(81)$ & $-0.2109(29)$ & $-0.2043(25)$ & $0.8815(63)$ \\
 & $0.121099$  & $0.875(12)$ & $-0.2142(66)$ &  $0.8692(70)$& $-0.2222(14)$ & $-0.2103(24)$ &  $0.8988(33)$\\
 \hline
$5.65$ & $0.122005$  & $0.8738(65)$ & $-0.2145(25)$ & $0.8738(65)$ & $-0.2145(25)$ & $-0.2145(25)$ & $0.8738(65)$ \\
 & $0.122078$  & $0.8861(62)$ & $-0.2050(34)$ & $0.8812(54)$ & $-0.2124(26)$ & $-0.2030(26)$  & $0.9043(51)$ \\
 & $0.122130$  & $0.815(34)$ & $-0.192(14)$ &$0.811(17)$  & $-0.2104(76)$ & $-0.1989(95)$ &  $0.8677(96)$\\
 & $0.122167$  & $0.8609(84)$ & $-0.2078(78)$ & $0.8513(62)$ &$-0.2206(19)$  &  $-0.2008(41)$& $0.9034(44)$ \\
 & $0.122197$  &$0.868(71)$& $-0.197(19)$ & $0.821(72)$ & $-0.206(22)$& $-0.198(13)$ &  $0.913(58)$\\
 \hline
 $5.80$& $0.122810$  & $0.866(13)$ & $-0.2062(55)$ &  $0.866(13)$ & $-0.2062(55)$ & $-0.2062(55)$&  $0.866(13)$ \\
 &  $0.122880$ & $0.8543(55)$ & $-0.2059(31)$&$0.8503(53)$ &$-0.2062(54)$  & $-0.1994(34)$ &  $0.8835(51)$ \\
 & $0.122940$  & $0.848(11)$ & $-0.1963(57)$&$0.8399(74)$ & $-0.2155(55)$ & $-0.1943(27)$ &  $0.9043(50)$  \\
 \hline
 $5.95$& $0.123411$  & $0.8649(47)$ & $-0.2058(28)$ &$0.8696(54)$  & $-0.2019(51)$ & $-0.2082(41)$ & $0.8517(92)$ \\
 & $0.123460$  & $0.828(21)$  &$-0.197(15)$  &$0.828(21)$ & $-0.197(15)$ & $-0.197(15)$ & $0.828(21)$ \\
 &  $0.123523$&  $0.8522(70)$ & $-0.2041(38)$ & $0.8502(61)$ & $-0.2112(23)$ & $-0.2005(27)$ & $0.8897(47)$\\
\hline
\hline
\end{tabular}
\caption{\label{tab:tensor}Table of the bare results for the individual quark contributions to the overall tensor charge in the nucleon, $\Sigma$ and $\Xi$ baryons.}
\end{table*}
\end{ruledtabular}

\begin{ruledtabular}
\begin{table*}[h!]
\centering
\begin{tabular}{ll|llllll} 
$\beta$ & $\kappa_l$  & $g_{A_P}^u$ & $g_{A_P}^d$ & $g_{A_\Sigma}^u$ & $g_{A_\Sigma}^s$ & $g_{A_\Xi}^u$ &  $g_{A_\Xi}^s$\\
\hline
 $5.40$&$0.119930$  &$1.025(24)$ & $-0.360(22)$ & $1.025(24)$& $-0.360(22)$  & $-0.360(22)$ &  $1.025(24)$  \\
 & $0.119989$ & $1.024(26)$&$-0.344(20)$  &$1.018(21)$ & $-0.333(16)$  &$-0.324(18)$ & $ 1.0421(15)$  \\
 & $0.120048$ & $1.019(48)$ &$-0.322(32)$ & $0.982(39)$ &  $-0.338(21)$&$-0.312(15)$ &  $1.031(18)$  \\
 & $0.120084$ & $1.028(35)$ & $-0.334(35)$ &$0.983(53)$ & $-0.340(10)$ &$-0.316(12)$ & $1.065(10)$ \\
 \hline
 $5.50$& $0.120900$ & $1.002(36)$ & $-0.306(58)$ & $1.002(36)$ & $-0.306(58)$ & $-0.306(58)$ & $1.002(36)$ \\
 & $0.121040$  & $0.959(74)$ & $-0.306(58)$ & $ 0.943(49)$ & $-0.302(29)$ &$-0.315(30)$  & $1.035(24)$ \\
 & $0.121095$  & $1.014(52)$ & $-0.399(43)$ &$0.978(21)$  & $-0.3155(94)$ &  $-0.2930(90)$& $1.0445(72)$ \\
 \hline
$5.50$& $0.120950$  & $1.009(49)$ & $-0.298(24)$ & $1.009(49)$ & $-0.298(24)$ & $-0.298(24)$ & $1.009(49)$ \\
 & $0.121040$  &$0.835(91)$  & $-0.343(47)$ & $0.913(45)$ & $-0.309(20)$ & $-0.299(23)$ & $1.017(28)$ \\
 & $0.121099$  & $0.964(88)$ & $-0.341(41)$ &$0.907(62)$  & $-0.292(30)$ & $-0.306(21)$ & $1.021(24)$ \\
  \hline
$5.65$ & $0.122005$  & $0.950(48)$ & $-0.294(34)$ &  $0.950(48)$ & $-0.294(34)$ & $-0.294(34)$ &  $0.950(48)$ \\
 & $0.122078$  &  $1.045(70)$& $-0.322(26)$ & $1.046(57)$ & $-0.317(19)$ & $-0.295(18)$ &$1.044(41)$  \\
 & $0.122130$  &  $ 0.973(91)$& $ -0.319(31)$ & $0.979(91)$ & $ -0.330(17)$ & $-0.287(18)$ & $1.070(24)$ \\
 & $0.122167$  & $ 1.113(93)$ & $-0.287(59)$ &  $ 1.024(42)$&  $-0.310(17)$& $-0.282(13)$ & $1.064(14)$ \\
 & $0.122197$  & $1.049(59)$ & $-0.335(76)$ &  $ 1.020(42)$& $-0.316(18)$ & $-0.257(22)$ & $1.107(16)$ \\
 \hline
 $5.80$& $0.122810$  & $0.966(63)$ & $-0.223(35)$ & $0.966(63)$  & $-0.223(35)$ & $-0.223(35)$ & $0.966(63)$  \\
 &  $0.122880$ & $1.042(56)$ & $-0.336(25)$ & $1.032(46)$ & $-0.333(15)$ & $-0.309(61)$ &  $1.041(21)$\\
 & $0.122940$  & $1.028(71)$ & $-0.326(35)$ & $0.987(86)$ & $-0.295(38)$ & $-0.296(25)$ &  $1.028(35)$\\
 \hline
 $5.95$& $0.123411$  & $0.975(33)$ & $-0.294(18)$ & $0.967(40)$ & $-0.270(38)$ & $-0.308(32)$ & $0.941(56)$ \\
 & $0.123460$  & $0.992(41)$ & $-0.308(22)$ & $0.992(41)$ & $-0.308(22)$ &$-0.308(22)$  & $0.992(41)$ \\
 &  $0.123523$&  $1.035(94)$ & $-0.364(62)$ & $1.004(63)$ & $-0.340(34)$ & $-0.302(36)$ & $0.960(56)$\\
\end{tabular}
\caption{\label{tab:axial}Table of the bare results for the individual quark contributions to the overall axial charge in the nucleon, $\Sigma$ and $\Xi$ baryons.}
\end{table*}
\end{ruledtabular}

\begin{ruledtabular}
\begin{table*}[h!]
\centering
\begin{tabular}{ll|llllll} 
$\beta$ & $\kappa_l$  & $g_{S_P}^u$ & $g_{S_P}^d$ & $g_{S_\Sigma}^u$ & $g_{S_\Sigma}^s$ & $g_{S_\Xi}^u$ &  $g_{S_\Xi}^s$\\
\hline
 $5.40$&$0.119930$  &$4.34(11)$ &$2.854(77)$  & $4.34(11)$& $2.854(77)$ &$2.854(77)$ & $4.34(11)$   \\
 & $0.119989$ &$4.32(14)$ & $2.987(92)$ & $4.03(12)$& $2.695(46)$ & $2.683(71)$&  $4.181(66)$ \\
 & $0.120048$ & $4.44(41)$ &$3.28(18)$ &  $3.98(15)$& $2.467(91)$ &$3.01(22)$ &   $4.43(13)$ \\
 & $0.120084$ & $4.29(64)$ &$2.62(47)$  & $4.04(20)$&  $2.702(42)$& $2.633(90)$&  $4.282(50)$\\
 \hline
 $5.50$& $0.120900$ & $4.26(10)$ & $2.766(74)$ & $4.26(10)$  & $2.766(74)$ & $2.766(74)$ & $4.26(10)$  \\
 & $0.121040$ & $4.93(43)$ & $3.41(24)$ & $4.35(25)$ & $2.642(60)$  & $2.73(10)$ & $4.201(70)$ \\
 & $0.121095$  & $5.60(29)$ &$4.13(23)$  & $4.10(19)$ & $ 2.514(34)$ & $2.474(93)$ & $4.028(39)$ \\
 \hline
$5.50$& $0.120950$  & $4.209(16)$ & $2.81(11)$ & $4.209(16)$ & $2.81(11)$  & $2.81(11)$  &$4.209(16)$  \\
 & $0.121040$  & $5.39(36)$ & $3.83(29)$ & $4.31(21)$ &$2.867(59)$  & $2.77(14)$ & $4.388(87)$ \\
 & $0.121099$  & $5.50(52)$ &$4.46(41)$  & $5.44(59)$ & $2.937(68)$ & $3.09(26)$ & $4.611(94)$ \\
 \hline
$5.65$ & $0.122005$  &$4.83(23)$  &$3.15(13)$  &  $4.83(23)$&$3.15(13)$  &  $3.15(13)$& $4.83(23)$ \\
 & $0.122078$  & $4.78(20)$ & $3.163(19)$ &$4.16(13)$  & $2.705(51)$ & $2.82(13)$ & $4.53(10)$ \\
 & $0.122130$  & $5.21(55)$ & $4.02(53)$ & $4.31(25)$&  $2.663(56)$& $2.55(14)$ & $4.242(63)$ \\
 & $0.122167$  & $5.73(39)$ & $4.15(22)$ & $4.01(23)$ & $2.759(46)$ & $2.61(13)$ & $4.344(70)$ \\
 & $0.122197$  & $6.34(59)$ & $4.36(40)$ & $4.04(38)$ & $2.755(75)$ & $2.50(21)$ &  $4.279(92)$\\
 \hline
 $5.80$& $0.122810$  & $4.47(27)$ & $2.89(16)$ & $4.47(27)$ & $2.89(16)$ & $2.89(16)$ & $4.47(27)$ \\
 &  $0.122880$ & $4.55(20)$ &$3.34(13)$  &$3.96(15)$  & $2.937(73)$ & $2.76(10)$ & $4.43(10)$ \\
 & $0.122940$  &$5.26(68)$  & $4.15(73)$ & $4.20(33)$ & $2.872(70)$ & $2.61(23)$ & $4.384(90)$ \\
 \hline
 $5.95$& $0.123411$ & $4.18(34)$ & $2.82(22)$ & $5.00(28)$ & $3.84(63)$ & $3.39(17)$ & $5.56(64)$ \\        
 & $0.123460$  & $5.21(27)$ & $3.38(17)$ & $5.21(27)$  & $3.38(17)$ & $3.38(17)$ & $5.21(27)$  \\
 &  $0.123523$& $4.84(28)$ & $3.57(23)$ & $4.24(24)$ & $2.901(90)$ & $2.93(18)$ & $4.29(10)$ \\
\end{tabular}
\caption{\label{tab:scalar}Table of the bare results for the individual quark contributions to the overall scalar charge in the nucleon, $\Sigma$ and $\Xi$ baryons.}
\end{table*}
\end{ruledtabular}
\clearpage
\section{Hyperon Matrix elements}
\label{appendix1}
Reconstruction of the hyperon matrix elements as shown to first order in Ref.~\cite{Bickerton_2019} and given to second order here:
\begin{eqnarray}
\bra{\Sigma^{+}}\bar{u}\Gamma u\ket{\Sigma^{+}}&=&2\sqrt{2}f + (-2\sqrt{2}s_1+\sqrt{6}s_2)\delta m_l + \sqrt{2}f_3\delta m_l^2, \nonumber\\
 \bra{\Sigma^{+}}\bar{s}\Gamma s\ket{\Sigma^{+}}&=&\sqrt{2}(f-\sqrt{3}d)+\Big(-\sqrt{\frac{3}{2}}r_1-3\sqrt{2}r_3-\sqrt{2}s_1\nonumber \\&+&\sqrt{\frac{3}{2}}s_2\Big)\delta m_l+\Big(-\sqrt{\frac{3}{2}}d_2+\frac{1}{\sqrt{2}}f_3\Big)\delta m_l^2, \nonumber\\
 \end{eqnarray} 
 \begin{eqnarray}
\bra{\Xi^{0}}\bar{u}\Gamma u\ket{\Xi^{0}}&=&\sqrt{2}(f-\sqrt{3}d)+(2\sqrt{2}r_3+2\sqrt{2}s_1)\delta m_l\nonumber \\&+&\Big(-\sqrt{\frac{3}{2}}d_4+\frac{1}{\sqrt{2}}f_2\Big)\delta m_l^2,\nonumber\\
\bra{\Xi^{0}}\bar{s}\Gamma s\ket{\Xi^{0}}&=&2\sqrt{2}f+\Big(-\sqrt{\frac{3}{2}}r_1+\sqrt{2}r_3+\sqrt{2}s_1\nonumber\\&-&\sqrt{\frac{3}{2}}s_2\Big)\delta m_l+\Big(\frac{\sqrt{3}}{2\sqrt{2}}d_1-\frac{\sqrt{3}}{2\sqrt{2}}d_4\nonumber\\&+& \frac{3}{2\sqrt{2}}f_1+\frac{1}{2\sqrt{2}}f_2\Big)\delta m_l^2.\nonumber
\end{eqnarray}
\bibliography{apssamp}

\end{document}